\RequirePackage{fix-cm}
\documentclass[twocolumn,epjc3]{svjour3}

\smartqed                        
\RequirePackage{graphicx}        
\RequirePackage{amsmath,amssymb}
\usepackage{float}
\RequirePackage{bm}              
\RequirePackage{booktabs,multirow,makecell,tabularx}
\RequirePackage{enumitem}
\RequirePackage[numbers,sort&compress]{natbib}
\RequirePackage{xcolor}
\RequirePackage[colorlinks, citecolor=blue,
                linkcolor=blue, urlcolor=blue]{hyperref}
\usepackage[T1]{fontenc}
\usepackage[utf8]{inputenc}
\usepackage[english]{babel}
\usepackage{microtype}

\usepackage{dblfloatfix}

\journalname{Eur. Phys. J. C}

\begin{document}
\title{Probing the Bounce Energy Scale in Bouncing Cosmologies\\
       with Pulsar Timing Arrays}

\author{Junrong~Lai\thanksref{addr1}
        \and
        Changhong~Li\thanksref{e1,addr1}}

\thankstext{e1}{e‑mail: \href{mailto:changhongli@ynu.edu.cn}{changhongli@ynu.edu.cn}}

\institute{Department of Astronomy, \\
           Key Laboratory of Astroparticle Physics of Yunnan Province, \\
           School of Physics and Astronomy, \\
           Yunnan University, No.\,2 Cuihu North Road, Kunming 650091, China
           \label{addr1}
          }

\date{\today}

\maketitle
\begin{abstract}
In this work, we constrain the bounce energy scale \(\rho_{s\downarrow}^{1/4}\) in a generic bouncing‐cosmology framework using the probable nanohertz stochastic gravitational‐wave signal from NANOGrav 15‐yr, EPTA DR2, PPTA DR3, and IPTA DR2. A full Bayesian fit to the analytic stochastic‐GW spectrum, for the first time, reveals two distinct posterior branches in \(\bigl(\rho_{s\downarrow}^{1/4},\,w_{1}\bigr)\): one near \(w_{1}\approx0.3\) and another at \(w_{1}\gg1\), where \(w_{1}\) is the contraction‐phase equation of state. We find that current PTA data strongly favor the bounce hypothesis over six conventional SGWB sources, while yielding smaller Bayes factors than its parent dual inflation–bounce scenario as the bounce scale is explicit. Both branches imply \(\rho_{s\downarrow}^{1/4}>M_{\rm pl}\), indicating a trans‐Planckian probe. In particular, the \(w_{1}\gg1\) branch violates the dominant energy condition, motivating novel early‐Universe physics. These results, confirmed by an accelerated a machine learning approach, offer a rare observational window into UV completions of cosmology.

\end{abstract}

\section{Introduction}

Currently, multiple pulsar timing array (PTA) collaborations—including NANOGrav~\cite{NANOGrav:2020spf,NANOGrav:2023gor}, EPTA~\cite{EPTA:2021crs,EPTA:2023fyk}, PPTA~\cite{Goncharov:2021oub,Reardon:2023gzh}, IPTA~\cite{Antoniadis:2022pcn}, and CPTA~\cite{Xu:2023wog}—report a common-spectrum process with Hellings–Downs correlations detected at up to \(3-4\sigma\) significance, which may be the first hint of a nanohertz stochastic gravitational wave background (SGWB).  This breakthrough opens a unique observational window onto both astrophysical and cosmological gravitational wave (GW) sources~\cite{NANOGrav:2023hvm,EPTA:2023xxk,Figueroa:2023zhu,Bian:2023dnv,Ellis:2023oxs,Caprini:2018mtu}, including supermassive–black‐hole binaries~\cite{Sesana:2013wja,Kelley:2017lek,Chen:2018znx,Burke-Spolaor:2018bvk,Calza:2024qxn}, inflationary GWs~\cite{Zhao:2013bba,Guzzetti:2016mkm,Vagnozzi:2020gtf}, cosmic strings~\cite{Siemens:2006yp,Cui:2018rwi,Gouttenoire:2019kij}, domain walls~\cite{Hiramatsu:2013qaa,Ferreira:2022zzo,Bian:2022qbh}, first‐order phase transitions~\cite{Hindmarsh:2013xza,Hindmarsh:2015qta,Caprini:2015zlo,Hindmarsh:2017gnf,Gouttenoire:2023bqy,Salvio:2023ynn}, scalar‐induced GWs~\cite{Ananda:2006af,Baumann:2007zm,Kohri:2018awv}, and more exotic early‐Universe scenarios such as audible axions~\cite{Machado:2019xuc,Machado:2018nqk}, kination domination~\cite{Co:2021lkc,Oikonomou:2023qfz}, preheating~\cite{Bethke:2013aba,Adshead:2019igv}, and particle‐production–driven signals~\cite{Dimastrogiovanni:2016fuu,DAmico:2021fhz}.  Together, these results inaugurate a powerful new probe of fundamental physics in cosmology and gravity.  

Remarkably, this nanohertz SGWB signal also provides a novel test of bouncing cosmologies~\cite{Li:2024oru} (see review in~\cite{Caprini:2018mtu} and recent developments in~\cite{Zhu:2023lbf, Papanikolaou:2024fzf, Ben-Dayan:2024aec, Qiu:2024sdd}).  Unlike standard inflationary scenarios—which posit a singular origin~\cite{Guth:1980zm,Starobinsky:1980te,Sato:1980yn,Linde:1981mu,Albrecht:1982wi,Mukhanov:1990me,Borde:1993xh,Borde:2001nh} (cf.\ recent discussions in~\cite{Lesnefsky:2022fen,Easson:2024uxe})—bouncing models avoid the singularity by passing through a finite, nonzero minimum scale factor during a contraction–expansion transition~\cite{Khoury:2001wf,Gasperini:2002bn,Creminelli:2006xe,Peter:2006hx,Cai:2007qw,Cai:2008qw,Saidov:2010wx,Li:2011nj,Cai:2011tc,Easson:2011zy,Bhattacharya:2013ut,Qiu:2015nha,Cai:2016hea,Barrow:2017yqt,deHaro:2017yll,Ijjas:2018qbo,Boruah:2018pvq,Nojiri:2019yzg,Silva:2015qna,Silva:2020bnn,Silva:2023ieb,Nayeri:2005ck,Brandenberger:2006xi,Fischler:1998st,Cai:2009rd,Li:2014era,Cheung:2014nxi,Li:2014cba,Li:2015egy,Li:2020nah} (see reviews in~\cite{Novello:2008ra,Brandenberger:2016vhg,Nojiri:2017ncd,Odintsov:2023weg}).  Bayesian comparisons between generic inflation and bounce scenarios yield Bayes factors near unity—evidence for an inflation–bounce duality in SGWB data~\cite{Li:2024oru,Li:2013bha} (see Eq.~\eqref{eq:bfzz} and Table~\ref{tab:bfint}).  

\begin{figure*}[htbp]
\centering 
\includegraphics[width=1\textwidth]{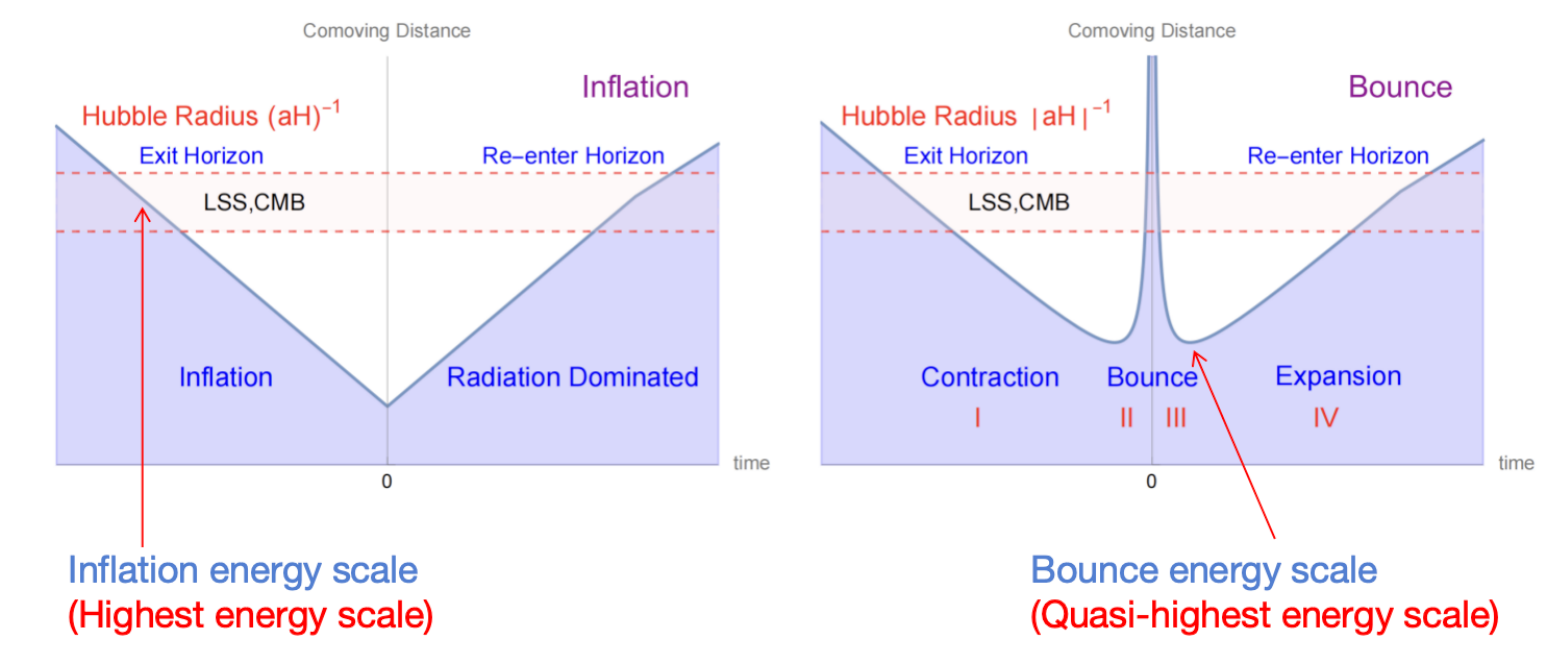}
\caption{\label{fig:IESBES}
Evolution of the comoving Hubble radius $|aH|^{-1}$ in (left) inflationary and (right) bouncing cosmologies~\cite{Cheung:2016vze,Li:2024dce,Li:2025ilc}.  In the bounce the horizon shrinks in Phases I and III and grows in Phases II and IV; primordial GWs exit in I, reenter in II, exit again in III, and finally reenter in IV, producing a double horizon–crossing history~\cite{Li:2024dce}.  Inflation maintains an almost constant “inflationary” energy scale, tightly constrained by current GW limits~\cite{Caprini:2018mtu}.  By contrast, bouncing models attain their maximum (“bounce”) energy scale at the central bounce~\cite{Li:2024dce,Li:2025ilc}, which we constrain here using PTA data.}
\end{figure*}

\begin{figure*}[htbp]
\centering 
\includegraphics[width=0.7\textwidth]{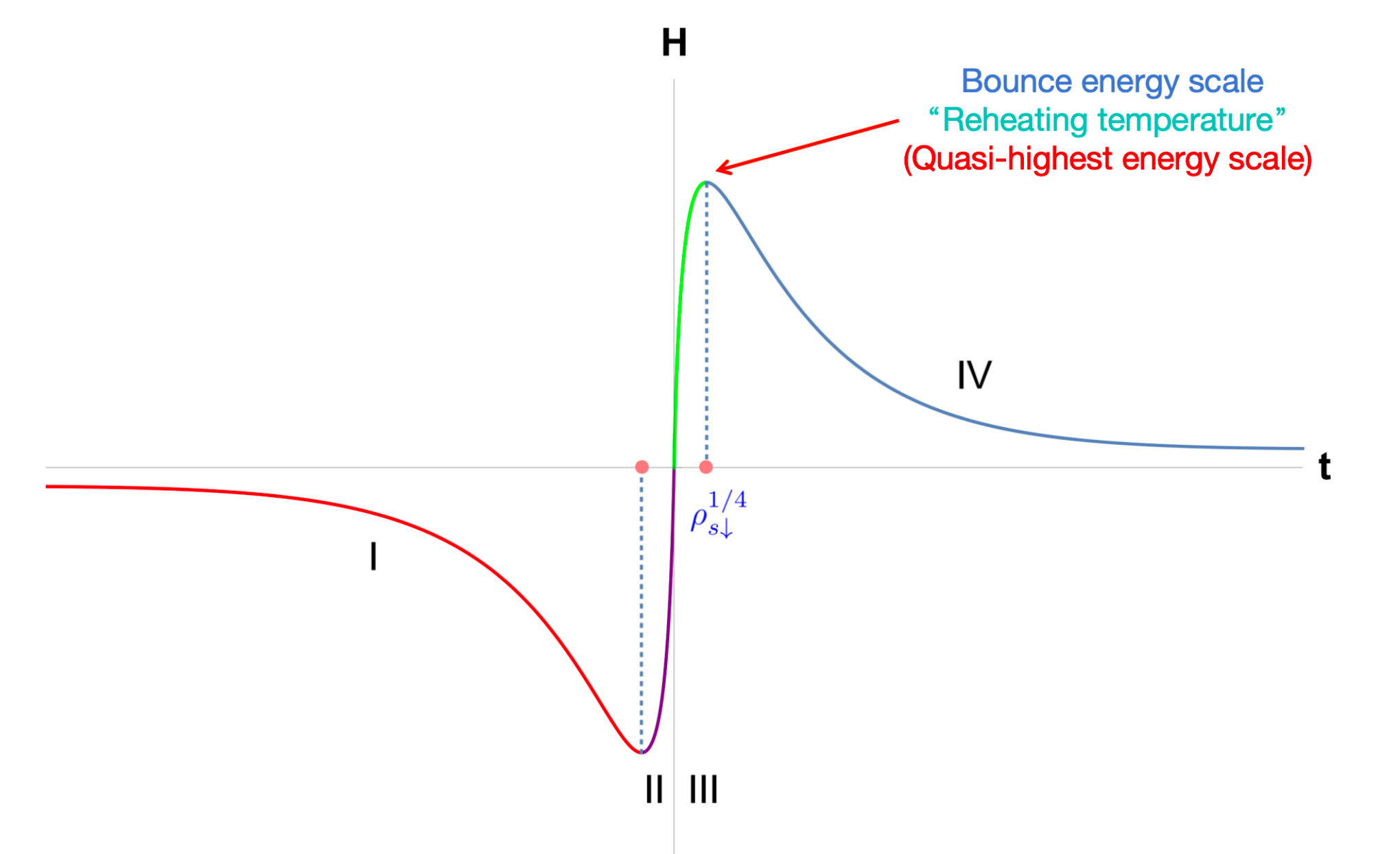}
\caption{\label{fig:Hbounce}
Evolution of the Hubble parameter \(H\) through the four phases of a generic non-singular bounce: contracting collapse (Phase I, \(H<0\), red), bounce transition (Phases II–III, \(H\) crosses zero, purple and green respectively), and post-bounce radiation era (Phase IV, \(H>0\), blue).  The peak value of \(|H|\) at the Phase III–IV boundary sets the bounce energy scale, here denoted as the quasi-highest scale \(\rho_{s\downarrow}^{1/4}\).}
\end{figure*}

Previous Bayesian treatments of bouncing models in PTA analyses~\cite{Li:2024oru} did not include the explicit bounce energy scale \(\rho_{s\downarrow}^{1/4}\), since the full evolution of primordial gravitational waves through the non-singular bounce was not computed.  As illustrated in the right panel of Fig.~\ref{fig:IESBES}, the scale factor of universe attains its minimum—and thus the quasi-highest energy scale—at the bounce point.  Incorporating \(\rho_{s\downarrow}^{1/4}\) therefore requires tracking the horizon crossings of PGWs across all four phases of the bounce, leading directly to the spectrum in Eq.~\eqref{eq:OmegaGWw1rhos}. In particular, Figure~\ref{fig:Hbounce} illustrates that \(\rho_{s\downarrow}^{1/4}\) corresponds to the maximum Hubble scale reached immediately after the non-singular bounce (onset of Phase IV).  Since additional microphysical effects could drive the instantaneous energy even higher at the bounce point, we refer to \(\rho_{s\downarrow}^{1/4}\) as the quasi-highest energy scale of the bouncing universe.

In a recent work~\cite{Li:2024dce} we established a general framework for the generation and evolution of primordial gravitational waves (PGWs) in non-singular bouncing cosmologies.  We partition the cosmic history into four phases—contracting collapse (I), bouncing contraction (II), bouncing expansion (III), and post-bounce radiation (IV)—as illustrated in Fig.~\ref{fig:IESBES}.  Employing the GW–propagation–matrix method of Ref.~\cite{Li:2024dce}, we derived an analytical expression for the PGW spectrum in bouncing cosmologies (c.f.\ Eq. (53) of Ref.~\cite{Li:2024dce}).  

Building on this, Ref.~\cite{Li:2025ilc} specialized to a concrete bounce by imposing three assumptions: (i) Phase IV is radiation dominated (\(w_4=1/3\)), as in the standard post-reheating Universe; (ii) the immediate pre- and post-bounce phases are governed by an exotic fluid with \(w_2=w_3\to-\infty\) (e.g.\ Quintom matter~\cite{Cai:2009zp}), ensuring a non-singular bounce; and (iii) the bounce is symmetric in conformal time, \(\eta_{s\downarrow}\equiv\eta_{1\downarrow}=\eta_{3\downarrow}\).  Under these conditions one obtains the closed-form SGWB spectrum in Eq.~\eqref{eq:OmegaGWw1rhos} (c.f.\ Eq. (24) of Ref.~\cite{Li:2025ilc}), which explicitly depends on the bounce energy scale \(\rho_{s\downarrow}^{1/4}\) (Eq.~\eqref{eq:rhosdd}).

In this work we employ the explicit SGWB spectrum for a non-singular bounce (Eq.~\eqref{eq:OmegaGWw1rhos}) to place the first direct PTA constraints on the bounce energy scale \(\rho_{s\downarrow}^{1/4}\) in a generic four-phase bouncing-cosmology framework.  Fitting to the combined NANOGrav 15-yr, EPTA DR2, PPTA DR3 and IPTA DR2 datasets~\cite{NANOGrav:2023gor,EPTA:2023fyk,Reardon:2023gzh,Antoniadis:2022pcn,ZenodoNG,NANOGrav:2023hde,ZenodoEPTA,EPTA:2023sfo,PPTADR3,Zic:2023gta,IPTADR2,Perera:2019sca} via full Bayesian inference, we uncover two distinct posterior branches in \((\rho_{s\downarrow}^{1/4},w_1)\): one at \(w_1\approx0.3\) and one at \(w_1\gg1\).  Recasting these bounds into the dual \((a_{s\downarrow},w_1)\) parametrization yields fully consistent results.  

We further compute Bayes factors against six standard SGWB sources—SMBHBs, inflationary GWs, cosmic strings, domain walls, first-order phase transitions and scalar-induced GWs—and find that the bounce model is strongly preferred in every case.  By contrast, comparison with the more generic “dual inflation–bounce” scenario \((w,r)\) yields smaller Bayes factors, demonstrating that making the bounce scale explicit tightens PTA constraints on this concrete realization.  

Furthermore, the two posterior regions uncovered by our PTA analysis highlight two theoretical frontiers in early-Universe physics:

\begin{enumerate}
  \item \textbf{Dominant-energy‐condition violation.} 
    The “right” branch (\(w_1\gg1\)) violates the dominant energy condition (Table~\ref{tab:nt_bestfit_by_dataset_range}), offering empirical impetus to explore novel pre-bounce physics such as ghost condensates, higher-derivative or modified-gravity operators, and extra-dimensional effects.
  \item \textbf{Trans-Planckian sensitivity.}  
    Both branches yield  
    \[
      \rho_{s\downarrow}^{1/4} \;>\; M_{\rm pl}\equiv G^{-1/2}\simeq1.22\times10^{16}\,\mathrm{TeV},
    \]  
    consistent with the lower bound found in Ref.~\cite{Li:2025ilc} (for the forecast constraints on $\rho_{s\downarrow}^{1/4}$ from high frequency gravitational wave using superconducting LC circuit and resonant cavities, see Ref.~\cite{Li:2025mxj}).  Thus, current PTAs are already probing trans-Planckian regimes in bouncing models.  Any proposed UV completion must respect these observational constraints, providing a rare window into physics beyond the Planck scale.
\end{enumerate}

\medskip

\noindent Finally, we have demonstrated (Sec.~\ref{sec:NFbasedML}) that a normalizing-flow–based machine-learning emulator reproduces our MCMC posterior with greatly reduced computational cost.  This approach is readily extendable to the expanded datasets anticipated from next-generation PTAs (e.g.\ SKA), enabling rapid and scalable Bayesian inference of bounce parameters.

\section{Stochastic Gravitational Waves for Bouncing Cosmologies}
In Ref.~\cite{Li:2025ilc}, considering a realistic bouncing cosmology that reheats to the standard radiation dominated era (\(w_{4}=1/3\)) after a sharp, symmetric bounce (\(w_{2}=w_{3}\to -\infty\), $\eta_{s\downarrow} \equiv \eta_{1\downarrow} = \eta_{3\downarrow}$), we derive the PGW‑induced stochastic gravitational‑wave background (SGWB) spectrum (c.f. Eq.(24) in Ref.~\cite{Li:2025ilc}), which encodes the bounce energy scale \(\rho_{s\downarrow}^{1/4}\):
\begin{equation}\label{eq:OmegaGWw1rhos}
\begin{split}
  \Omega_\mathrm{GW}(f)\,h^2
  &= \frac{h^2}{24}
    \left(\frac{f_{H_0}}{f_{m_\mathrm{pl}}}\right)^2
    \frac{C(w_1)}{(2\pi)^{-n_T(w_1)-1}}\\
  &\quad\times
    \left(\frac{f}{f_{H_0}}\right)^{n_T(w_1)}
    \left[\frac{\rho_{s\downarrow}^{1/4}}{(\rho_{c0}/\Omega_{\gamma0})^{1/4}}\right]^{4-n_T(w_1)}
    \mathcal{T}_\mathrm{eq}(f)\,.
\end{split}
\end{equation}
Here, $\rho_{s\downarrow}^{1/4}$ is defined at the beginning of Phase IV (the reheating point) with the assumption that Phase IV is standard post-reheating radiation-dominated era (c.f. Eq.(20) in Ref.~\cite{Li:2025ilc}), 
\begin{equation}\label{eq:rhosdd}
    \rho_{s\downarrow} \equiv \rho (\eta_s)= 3H^2(\eta_{s\downarrow}) m_\mathrm{pl}^2~,
\end{equation}
$C(w_1)$ is the amplitude factor~\cite{Li:2025ilc},
\begin{equation}
  C(w_1)=
  \begin{cases}
    \displaystyle
    \begin{aligned}
      \pi^{-1}\,4^{-1 + x(w_1)}\,\Gamma^2\!\bigl[x(w_1)\bigr] \\
      \qquad\times 
      \left[\frac{2(3w_1 - 1)}{3w_1 + 1}\right]^2
    \end{aligned}
    & -\tfrac13 \le w_1 < 1, \\[1ex]
    \displaystyle\\
    \pi^{-1}\,4^{-1 - x(w_1)}\,\Gamma^2\!\bigl[-x(w_1)\bigr]
    & w_1 \ge 1,
  \end{cases}
  \label{eq:Cw1}
\end{equation}

$n_T(w_1)$ is the tilt of the primordial tensor spectrum, depending on $w_1$~\cite{Li:2025ilc},  
\begin{equation}
  n_{T}\equiv 
  \frac{\mathrm{d}\ln\mathcal{P}_{h}(\eta_{k})}{\mathrm{d}\ln k}
  = 
  \begin{cases}
    3-2x(w_1), & -\tfrac13 \le w_1 < 1.\\
    3+2x(w_1), & w_1 \ge 1,
  \end{cases},
  \label{eq:nt}
\end{equation}
where \(w_i\) denotes the equation‑of‑state parameter in phase \(i\), \(x(w_1)\equiv \tfrac{3(1-w_1)}{2(3w_1+1)}\) and 
\(\Gamma(z)=\int_{0}^{\infty} t^{z-1}e^{-t}\,\mathrm{d}t\) is the Gamma function. The present‑day critical density is  
\(\rho_{c0}=3H_{0}^{2}m_{\mathrm{pl}}^{2}=8.098\,h^{2}\times10^{-11}\,\mathrm{eV}^{4}\)  
with the reduced Hubble parameter \(h=0.677\).  
Reference frequencies  
\(f_{H_{0}}=2.2\times10^{-18}\,\mathrm{Hz}\) and  
\(f_{m_{\mathrm{pl}}}=3.7\times10^{42}\,\mathrm{Hz}\)  
correspond to \(H_{0}\) and the reduced Planck mass  
\(m_{\mathrm{pl}}\equiv (\sqrt{8\pi G})^{-1}\), respectively, in natural units (\(\hbar=c=1\)).  
The radiation density fraction today is  
\(\Omega_{\gamma0}h^{2}=2.474\times10^{-5}\). The transfer function
\begin{equation}
  \mathcal{T}_{\mathrm{eq}}(f)=
  \biggl[1+\frac{9}{32}\Bigl(\frac{f_{\mathrm{eq}}}{f}\Bigr)^{2}\biggr],
\end{equation}
captures the evolution from radiation to matter domination, with  
\(f_{\mathrm{eq}}=2.01\times10^{-17}\,\mathrm{Hz}\) marking matter–radiation equality. For the frequency band accessible to present gravitational‑wave experiments—CMB/BICEP, pulsar timing arrays, and Advanced LIGO–Virgo—the equality‑era transfer function satisfies \(\mathcal{T}_{\mathrm{eq}}(f)\simeq 1\) because \(f\gg f_{\mathrm{eq}}\).  
Accordingly, the stochastic gravitational‑wave background observed today inherits the primordial tensor tilt \(n_T(w_1)\).

\begin{figure*}[htbp]
\centering 
\includegraphics[width=1\textwidth]{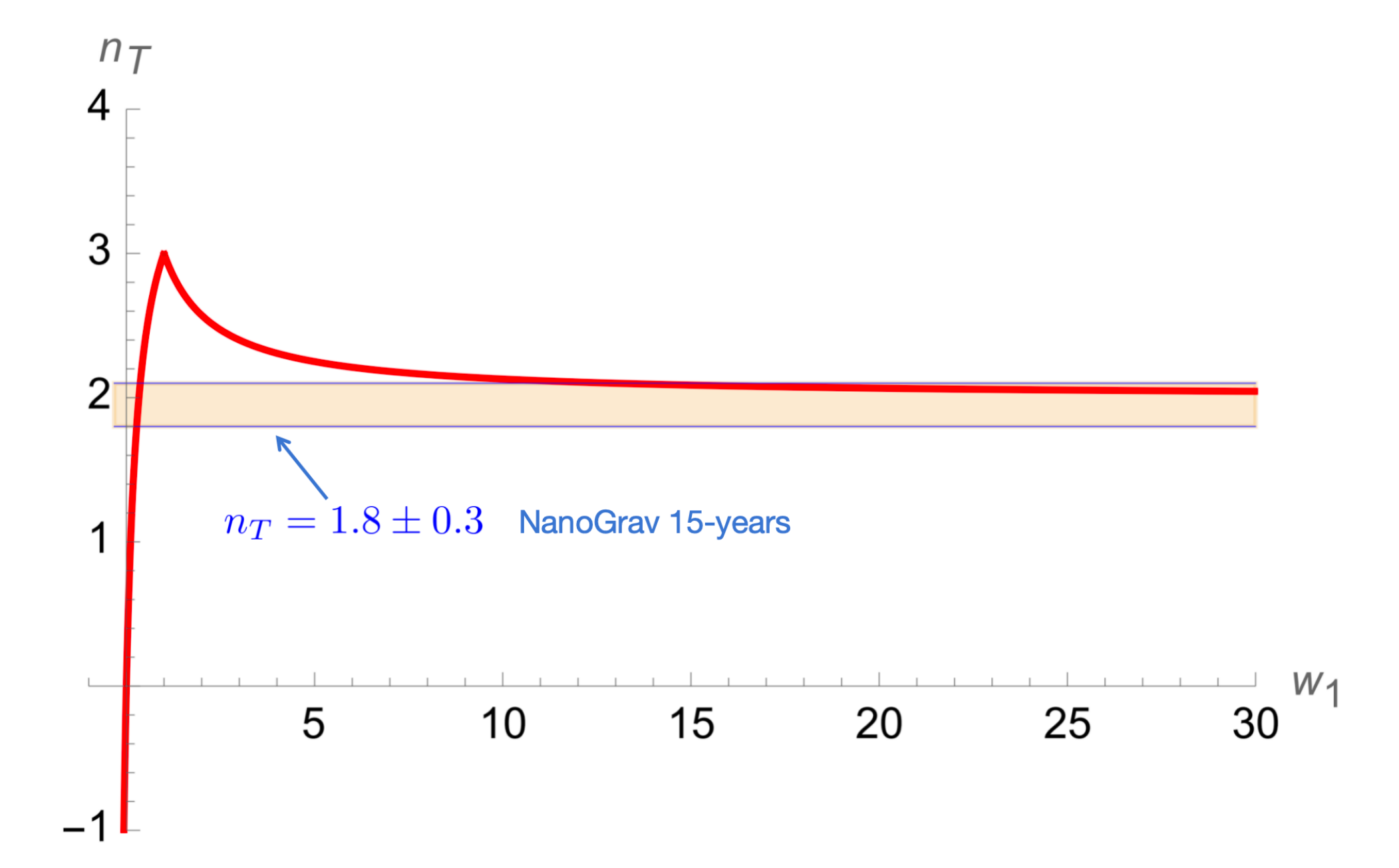}
\caption{\label{fig:ntvsw1} Illustration of the spectral index of the SGWB in bouncing cosmologies by using Eq.~(\ref{eq:nt}): $n_T$ as a function of $ w_1 $. The SGWB spectrum for big bounce cosmology is red-tilted ($ n_T < 0 $) for  $ w_1 < 0 $, blue-tilted ($ n_T > 0 $)  for $ w_1 > 0 $, and scale-invariant ($ n_T = 0 $) for $ w_1 = 0 $.}
\end{figure*}

Figure~\ref{fig:ntvsw1} shows the tensor spectral tilt \(n_T(w_1)\) (red curve) obtained from Eq.~\eqref{eq:nt}.  
The stochastic background is red‑tilted (\(n_T<0\)) for \(w_1<0\), blue‑tilted (\(n_T>0\)) for \(w_1>0\), and strictly scale‑invariant (\(n_T=0\)) at \(w_1=0\) (a matter‑dominated bounce).  
Consequently, low‑frequency detectors (CMB/BICEP)—most sensitive to red tilts—set the strongest bounds in the \(w_1<0\) regime, whereas high‑frequency instruments (aLIGO+aVirgo) constrain more tightly when \(w_1>0\).

Current pulsar‑timing‑array (PTA) measurements prefer a blue-tilted spectrum: the NANOGrav 15‑year data yield \(n_T = 1.8 \pm 0.3\) (68\% C.L.)~\cite{Vagnozzi:2023lwo}, shown as the yellow band in Fig.~\ref{fig:ntvsw1}.  
Its intersection with the red curve selects two allowed intervals, 
\begin{equation}\label{eq:twobranch}
\begin{cases}
    \mathbf{Branch 1:} & 0.20 < w_1 < 0.37; \\ 
    \mathbf{Branch 2:} & w_1 > 13, 
\end{cases}
\end{equation}
in which bouncing cosmologies can reproduce the observed blue tilt ($n_T(w_1)=1.8\pm 0.3$).  
Matching the amplitude of the signal, however, requires different bounce energy scales \(\rho_{s\downarrow}^{1/4}\) for different values of \(n_T(w_1)\). In the next section we perform a Bayesian analysis using the full PTA data set (NANOGrav 15-year, EPTA DR2, PPTA DR3 and IPTA DR2), which simultaneously constrains the amplitude and tilt of the spectrum, to determine the joint parameter region \((w_1,\rho_{s\downarrow}^{1/4})\) compatible with current observations.

\section{Interpretation of Pulsar-Timing-Array Data in Bouncing Cosmology}
\label{sec:interpretation}
\subsection{Posterior distribution of bounce energy scale and physics interpretation}
\begin{figure*}[htbp]
\centering 
\includegraphics[width=0.7\textwidth]{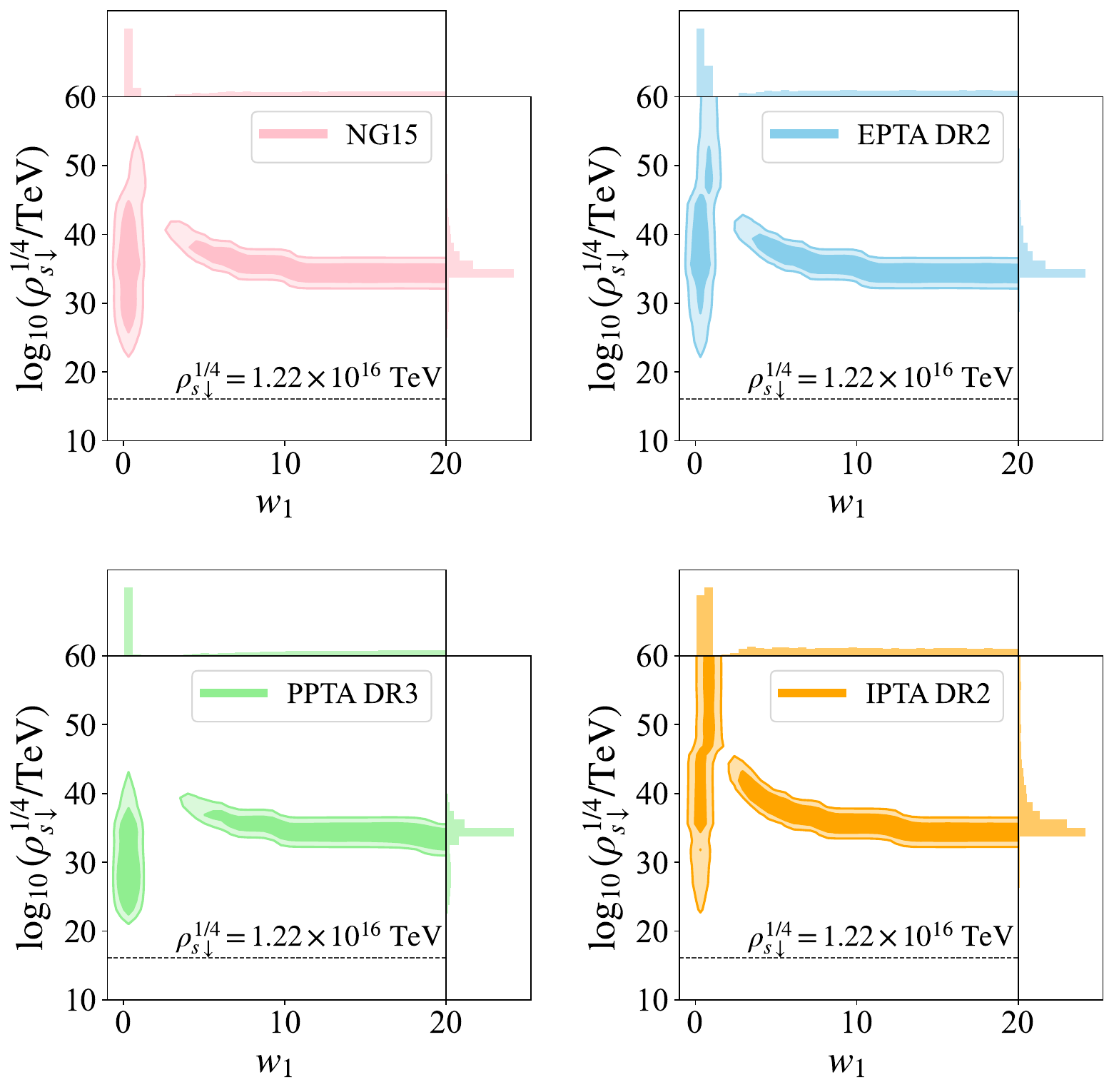}
\caption{\label{fig:2x2rhosw1}Two-dimensional posterior distribution in the \((\rho_{s\downarrow}^{1/4},w_1)\) plane, obtained by fitting Eq.~(\ref{eq:OmegaGWw1rhos}) to the combined PTA datasets (NANOGrav 15-yr, EPTA DR2, PPTA DR3, IPTA DR2).  The dashed horizontal line marks the  Planck scale \(M_{\rm pl}= G^{-1/2}\simeq1.22\times10^{16}\,\mathrm{TeV}\).}
\end{figure*}

\begin{figure*}[htbp]
\centering 
\includegraphics[width=0.7\textwidth]{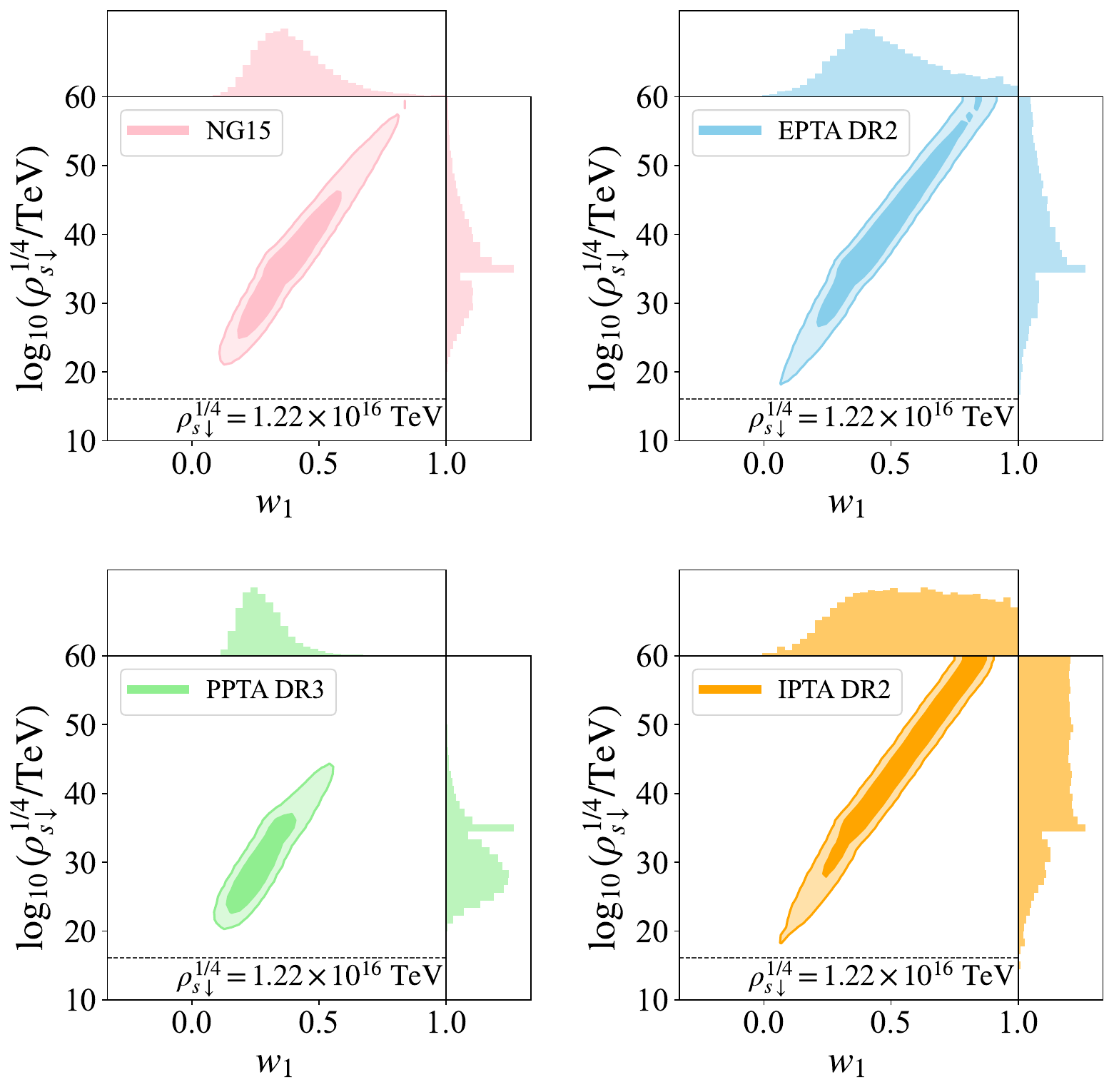}
\caption{\label{fig:2x2matrix131}High-resolution view of the “left” posterior branch in Fig.~\ref{fig:2x2rhosw1}, highlighting the region satisfying all energy conditions.}
\end{figure*}

\begin{figure*}[htbp]
\centering 
\includegraphics[width=1\textwidth]{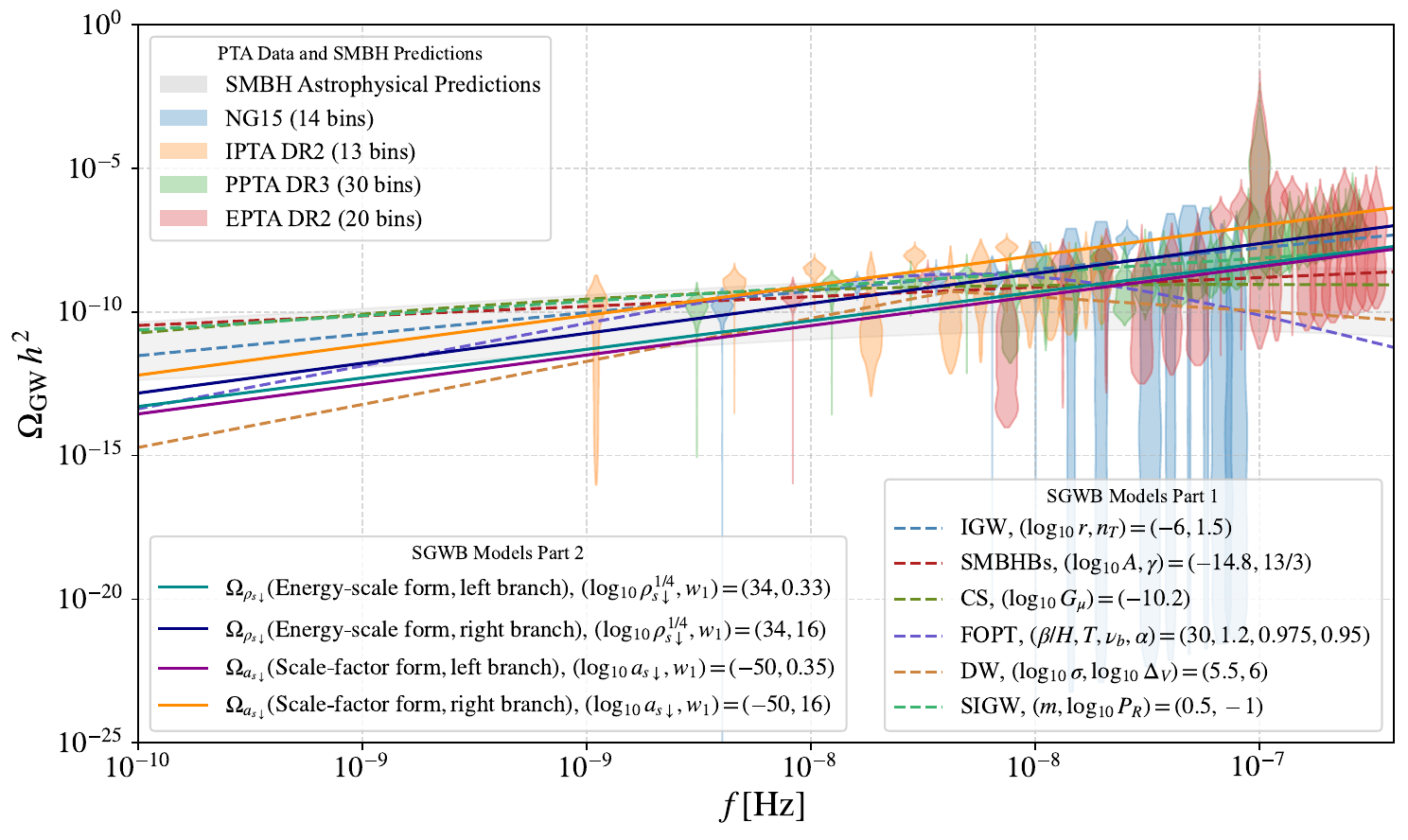}
\caption{\label{fig:Violinplot}
Violin plots of posterior SGWB amplitudes in each PTA frequency bin for two representative parameter sets—one from each of the \(1\sigma\) posterior branches in the \((\rho_{s\downarrow}^{1/4},w_1)\) plane (solid green and blue; Fig.~\ref{fig:2x2rhosw1}), with examples in the \((a_{s\downarrow},w_1)\) parametrization shown in solid purple and orange (Fig.~\ref{fig:2x2asw1}).  Dashed line correspond to six conventional SGWB models (Sec.~\ref{sec:SGWBmodels}).  The gray band shows the SMBHB astrophysical prediction~\cite{Rosado:2015epa}. Violins are built from the combined PTA datasets~\cite{NANOGrav:2023gor,EPTA:2023fyk,Reardon:2023gzh,Antoniadis:2022pcn,ZenodoNG,NANOGrav:2023hde,ZenodoEPTA,EPTA:2023sfo,PPTADR3,Zic:2023gta,IPTADR2,Perera:2019sca} using the method of Ref.~\cite{Gouttenoire:2023bqy}.}
\end{figure*}

To determine the region in the joint parameter space \((\rho_{s\downarrow}^{1/4}, w_1)\) of bouncing cosmologies that is compatible with current PTA measurements, we carry out a Bayesian fit of Eq.~(\ref{eq:OmegaGWw1rhos}) to the full PTA dataset ( NANOGrav 15-year, EPTA DR2, PPTA DR3 and IPTA DR2)~\cite{NANOGrav:2023gor,EPTA:2023fyk,Reardon:2023gzh,Antoniadis:2022pcn,ZenodoNG,NANOGrav:2023hde,ZenodoEPTA, EPTA:2023sfo,PPTADR3,Zic:2023gta, IPTADR2,Perera:2019sca}, employing \texttt{Ceffyl}~\cite{Lamb:2023jls}.  In Fig.~\ref{fig:2x2rhosw1} we present the resulting two-dimensional posterior distribution in the \((\rho_{s\downarrow}^{1/4}, w_1)\) plane for the bouncing scenario. Fig.~\ref{fig:2x2rhosw1} shows two distinct posterior branches, one near \(w_1 \approx 0.3\) and the other at \(w_1 \gg 1\), consistent with our analytic behavior of \(n(w_1)\) in Eq.~\eqref{eq:twobranch}.  The small offset between the posterior peaks and the analytic prediction ($0.20 < w_1 < 0.37$ and $w_1 > 13$) arises from the statistical uncertainties in the PTA data.  More precisely, Fig.~\ref{fig:2x2rhosw1} shows that on the “left” branch the inferred bounce scale \(\rho_{s\downarrow}^{1/4}\) grows with \(w_1\) (see the high-resolution inset in Fig.~\ref{fig:2x2matrix131}), reflecting the monotonic rise of \(n_T(w_1)\) in that region, whereas on the “right” branch \(\rho_{s\downarrow}^{1/4}\) asymptotes to a constant value as \(n_T(w_1)\to2\) for \(w_1\gg1\) (the amplitude factor \(C(w_1)\) must also be accounted for in a more detailed description). Additionally, in Fig.~\ref{fig:Violinplot}, we show violin plots for two representative points—one from each of the \(1\sigma\) posterior branches in the \((\rho_{s\downarrow}^{1/4},w_1)\) plane (Fig.~\ref{fig:2x2rhosw1})—alongside examples in the \((a_{s\downarrow},w_1)\) parametrization (Fig.~\ref{fig:2x2asw1}) and for six conventional SGWB source models (Sec.~\ref{sec:SGWBmodels}).  These plots confirm that both bounce-model branches identified in this work can explain the nanohertz SGWB observed by current PTA datasets.

Furthermore, from a theoretical perspective the “left” branch at \(w_1\approx0.3\) satisfies all standard energy conditions—null (\(w_1\ge-1\)), weak (\(\rho_{s\downarrow}\ge0\) and \(w_1\ge-1\)), strong (\(w_1\ge-\tfrac{1}{3}\)), and dominant (\(\rho_{s\downarrow}\ge0\) and \(-1\le w_1\le1\))—as confirmed by the high-resolution inset in Fig.~\ref{fig:2x2matrix131}. By contrast, the “right” branch at \(w_1\gg1\) violates the dominant energy condition (Table~\ref{tab:nt_bestfit_by_dataset_range}).  Violations of energy conditions have attracted considerable interest as effective descriptions of novel physics—e.g.\ ghost condensates, higher-derivative or modified-gravity operators, and extra-dimensional effects—operating during the very early Universe (inflation or the contracting phase of a bounce) (see, e.g., Refs.~\cite{Li:2024oru,Cai:2023uhc,Ye:2023tpz,Lohakare:2022umj,Gouttenoire:2021jhk,Bhattacharya:2015nda,Cai:2013kja,Liu:2012fk,Cai:2009zp,Santos:2007pp,Santos:2007bs,Burgess:2003tz}).  Our PTA-based constraints therefore offer complementary empirical motivation for pursuing bounce models that violate the dominant energy condition.  

\begin{table*}[htbp]
\centering
\caption{Best‐fit values for $n_T$, $w_1$ and $\rho_{s\downarrow}^{1/4}$ determined in Bayesian inference}
\small
\begin{tabular}{llccc}
\toprule
\textbf{Dataset} & \textbf{$w_1$ range} & \textbf{Best‐fit $n_T$} & \textbf{Best‐fit $w_1$} & \textbf{Best‐fit $\log_{10}(\rho_s^{1/4}/\mathrm{TeV})$} \\
\midrule
\textbf{NG 15}     & $[-\frac{1}{3},1]$ & $1.871$ & $0.293$  & \text{31.317
} \\
\textbf{EPTA DR2}  & $[-\frac{1}{3},1]$ & $2.120$ & $0.376$  & \text{35.885} \\
\textbf{PPTA DR3}  & $[-\frac{1}{3},1]$ & $1.676$ & $0.240$  & \text{28.053} \\
\textbf{IPTA DR2}  & $[-\frac{1}{3},1]$ & $2.262$ & $0.434$  & \text{39.147} \\
\midrule
\textbf{NG 15}     & $[1,20]$           & $2.067$ & $19.497$ & \text{33.880} \\
\textbf{EPTA DR2}  & $[1,20]$           & $2.123$ & $10.514$ & \text{34.957} \\
\textbf{PPTA DR3}  & $[1,20]$           & $2.066$ & $19.770$ & \text{33.681} \\
\textbf{IPTA DR2}  & $[1,20]$           & $2.259$ & $4.820$  & \text{38.402} \\
\bottomrule
\end{tabular}
\label{tab:nt_bestfit_by_dataset_range}
\end{table*}

Another point of theoretical interest is that the best‐fit bounce energy scale exceeds the Planck scale (Table~\ref{tab:nt_bestfit_by_dataset_range}), 
\begin{equation}
    \rho_{s\downarrow}^{1/4} \;>\; M_{\rm pl}\equiv G^{-1/2}\simeq1.22\times10^{16}\,\mathrm{TeV},
\end{equation}
for both posterior branches. This finding is consistent with the PTA‐only lower bound derived in Ref.~\cite{Li:2025ilc}, where SGWB limits and forecasts—from PTAs, Planck/BICEP, aLIGO+aVirgo (O2), CMB-S4, IPTA (design), SKA, DECIGO, BBO, TianQin, Taiji, aLIGO+aVirgo+KAGRA (design), Cosmic Explorer, and Einstein Telescope—were used to constrain \(\rho_{s\downarrow}^{1/4}\) across multiple bands~\cite{Schmitz:2020syl,Annis:2022xgg,Bi:2023tib}.Hence, current pulsar-timing arrays are already probing trans-Planckian physics in bouncing-cosmology scenarios.  Any viable UV completion must therefore respect these PTA bounds, providing a rare observational window into physics beyond the Planck scale.  This situation echoes the “trans-Planckian problem” of inflation~\cite{Martin:2000xs,Brandenberger:2000wr,Brandenberger:2012aj,Kaloper:2002cs,Easther:2002xe,Bedroya:2019snp,Bedroya:2019tba,Benetti:2021uea, Bedroya:2022tbh} (see ~\cite{Bedroya:2024zta, Bedroya:2025ris} for recent development.), in which modes originating above \(m_{\rm pl}\) nonetheless leave imprints on low-energy observables, highlighting the need for new quantum-gravitational input to render predictions unambiguous.

\subsection{Bayes‐factor comparison with conventional SGWB sources}
Furthermore, as shown in Table~\ref{tab:rho_vs_other_1f}, we assess the preference for the bouncing-cosmology model against six standard SGWB sources (SMBHBs, IGW, CS, DW, FOPT, and SIGW; for model definitions see Sec.~\ref{sec:SGWBmodels}) by computing the Bayes factor  
\begin{equation}\label{eq:bfzz}
    \mathrm{BF}_{ij} \;=\;\frac{\mathcal{Z}_i}{\mathcal{Z}_j}\,,
\end{equation}
where \(\mathcal{Z}_i\) is the evidence for model \(M_i\).  For two models \(M_1\) and \(M_2\), \(\mathrm{BF}_{12}>1\) indicates that \(M_1\) is favored over \(M_2\) (Table~\ref{tab:bfint}).  Table~\ref{tab:rho_vs_other_1f} shows that every PTA dataset yields \(\mathrm{BF}>1\) for all comparisons, implying that the bouncing-cosmology parameterization \((\rho_{s\downarrow}^{1/4},w_1)\) is preferred over each of the six conventional SGWB source models.  
NANOGrav 15-yr delivers the strongest discrimination—decisive vs.\ CS (\(\mathrm{BF}=161.9\)) and very strong vs.\ SMBHBs (\(\mathrm{BF}=33.2\)), with strong support vs.\ DW (\(\mathrm{BF}=111.9\)), FOPT (\(\mathrm{BF}=21.5\)) and SIGW (\(\mathrm{BF}=22.9\)), and moderate preference vs.\ IGW (\(\mathrm{BF}=2.1\)).  PPTA DR3 similarly yields strong to moderate evidence vs.\ SMBHBs (\(\mathrm{BF}=7.3\)), DW (\(\mathrm{BF}=6.6\)), CS (\(\mathrm{BF}=5.4\)) and SIGW (\(\mathrm{BF}=4.0\)), but only weak preference vs.\ IGW (\(\mathrm{BF}=1.1\)) and FOPT (\(\mathrm{BF}=2.1\)).  EPTA DR2 shows strong preference only against DW (\(\mathrm{BF}=11.4\)) and weak-to-moderate against others (\(\mathrm{BF}=1.2\!-\!2.3\)), while IPTA DR2 gives moderate support vs.\ CS, DW and SIGW (\(\mathrm{BF}=3.6\!-\!6.5\)) and weak evidence vs.\ SMBHBs, IGW, FOPT (\(\mathrm{BF}=1.3\!-\!2.5\)).  Thus, the bounce spectrum consistently outperforms standard SGWB models, with the most decisive results from NG15 and PPTA DR3.  

\begin{table*}[htbp]
\centering
\caption{Bayes factors for the bouncing‐universe model \((\rho_{s\downarrow}^{1/4},w_1)\) versus six conventional SGWB source models.}
\small 
\hspace*{-1.cm}
\begin{tabular}{cccccccc}
\toprule
\textbf{Data Set} & \textbf{$(\rho_{s\downarrow}^{1/4},w_1)$/SMBHBs} & \textbf{$(\rho_{s\downarrow}^{1/4},w_1)$/IGW} & \textbf{$(\rho_{s\downarrow}^{1/4},w_1)$/CS} & \textbf{$(\rho_{s\downarrow}^{1/4},w_1)$/DW} & \textbf{$(\rho_{s\downarrow}^{1/4},w_1)$/FOPT} & \textbf{$(\rho_{s\downarrow}^{1/4},w_1)$/SIGW} \\ \hline
\textbf{NG15} & 33.2 ± 13.2 & 2.1 ± 0.7 & 161.9 ± 48.2 & 111.9 ± 27.8 & 21.5 ± 4.4 & 22.9 ± 4.7 \\
\textbf{EPTA DR2} & 1.5 ± 0.6 & 1.4 ± 0.5 & 1.2 ± 0.4 & 11.4 ± 2.9 & 1.2 ± 0.3 & 2.3 ± 0.5 \\ 
\textbf{PPTA DR3} & 7.3 ± 3.0 & 1.1 ± 0.4 & 5.4 ± 2.3 & 6.6 ± 2.9 & 2.1 ± 0.8 & 4.0 ± 1.4 \\
\textbf{IPTA DR2} & 1.8 ± 0.9 & 2.5 ± 1.1 & 3.6 ± 1.6 & 3.4 ± 1.8 & 1.3 ± 0.5 & 6.5 ± 3.0 \\ \bottomrule
\end{tabular}
\label{tab:rho_vs_other_1f}
\end{table*}


\begin{table}[htbp]
\centering
\caption{Interpretation of Bayes‐factor values for model comparison, following the Jeffreys scale \cite{Bian:2023dnv,Lai:2025xov}.}
\begin{tabular}{c c }
\toprule
$\mathrm{BF}_{ij}$ & Evidence Strength for $M_i$ vs $M_j$ \\
\midrule
1--3 & Weak \\
3--20 & Positive \\ 
20--150 & Strong \\
$\geq 150$ & Very strong \\
\bottomrule
\end{tabular}
\label{tab:bfint}
\end{table}

\subsection{Comparing with the dual scenario of inflation-bounce cosmologies}
\begin{figure}[htbp]
\centering 
\includegraphics[width=0.5\textwidth]{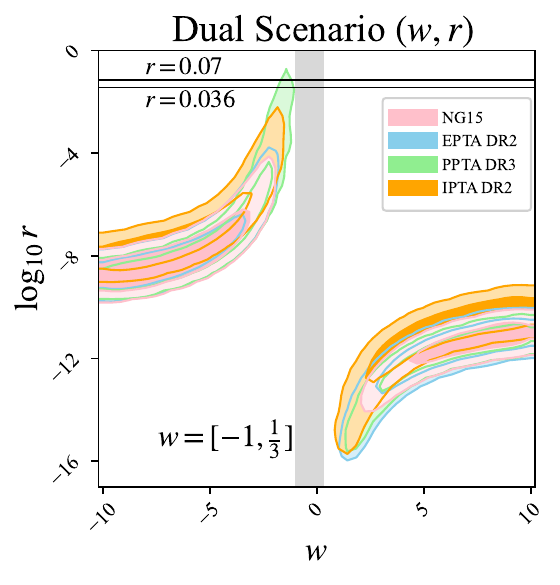}
\caption{\label{fig:dualwnt}Posterior distribution of dual scenario of inflation-bounce cosmologies $(w,r)$, obtained by fitting its SGWB spectrum to the same full PTA dataset, c.f. Figure 3 in Ref.~\cite{Li:2024oru}.}
\end{figure}

The dual scenario of inflation-bounce cosmologies \((w,r)\)~\cite{Li:2013bha}, obtained by extending Wands’ duality~\cite{Wands:1998yp, Finelli:2001sr, Boyle:2004gv, Raveendran:2023auh}, predicts both a nearly scale‐invariant CMB spectrum \((n_s-1\simeq0)\) and a blue‐tilted SGWB \((n_T=1.8\pm0.3)\) consistent with CMB/BICEP and PTA data~\cite{Li:2024oru} (Fig.~\ref{fig:dualwnt}).  Our bouncing‐universe model is a concrete realization of this scenario, with an explicit SGWB spectrum $\Omega_\mathrm{GW}(f)=\Omega_\mathrm{GW}\bigl(f,\rho_{s\downarrow}^{1/4}\bigr)$ that introduces the bounce energy scale \(\rho_{s\downarrow}^{1/4}\).  Because the form of \(\Omega_\mathrm{GW}(f,\rho_{s\downarrow}^{1/4})\) is fixed in this concrete realization, PTAs impose tighter constraints on the \((\rho_{s\downarrow}^{1/4},w_1)\) parameter space, yielding smaller Bayes factors relative to the more generic \((w,r)\) dual scenario.  Table~\ref{tab:rho_vs_other} presents the Bayes‐factor comparison between \((\rho_{s\downarrow}^{1/4},w_1)\) and \((w,r)\), confirming that PTA data favor the concrete bouncing‐universe realization while still supporting the overarching dual scenario.More specifically, Table~\ref{tab:rho_vs_other} shows that every Bayes factor for \((\rho_{s\downarrow}^{1/4},w_1)\) versus either branch—or the bounce‐only branch—of the dual scenario \((w,r)\) lies below unity, as anticipated.  

\begin{table*}[htbp]
\centering
\caption{Bayes factors for the bouncing‐universe model \((\rho_{s\downarrow}^{1/4},w_1)\) versus dual scenario of inflation-bounce cosmologies $(w,r)$~\cite{Li:2024oru, Li:2013bha}.}
\small 
\begin{tabular}{ccccc}
\toprule
\textbf{Data Set} & \textbf{$(\rho_{s\downarrow}^{1/4},w_1)$/Dual$(w,r)$} & \textbf{$(\rho_{s\downarrow}^{1/4},w_1)$/Bounce} & \textbf{$(a_{s\downarrow},w_1)$/Dual$(w,r)$} & \textbf{$(a_{s\downarrow},w_1)$/Bounce} \\ \hline
\textbf{NG15} & 0.4 $\pm$ 0.1 & 0.4 $\pm$ 0.2 & 0.3 $\pm$ 0.1 & 0.3 $\pm$ 0.1 \\
\textbf{EPTA DR2} & 0.3 $\pm$ 0.1 & 0.3 $\pm$ 0.1 & 0.3 $\pm$ 0.1 & 0.3 $\pm$ 0.1 \\
\textbf{PPTA DR3} & 0.3 $\pm$ 0.1 & 0.3 $\pm$ 0.1 & 0.4 $\pm$ 0.1 & 0.4 $\pm$ 0.2 \\
\textbf{IPTA DR2} & 0.4 $\pm$ 0.2 & 0.4 $\pm$ 0.2 & 0.2 $\pm$ 0.1 & 0.2 $\pm$ 0.1 \\ 
\bottomrule
\end{tabular}
\label{tab:rho_vs_other}
\end{table*}

\section{Posterior distribution of bounce scale factor}

In this section we recast the PTA bounds on the bounce energy scale \(\rho_{s\downarrow}^{1/4}\) into an equivalent constraint on the bounce scale factor \(a_{s\downarrow}\).  Here \(a_{s\downarrow}\) is defined as the quasi‐minimal size of the Universe at the bounce (the boundary between Phases III and IV), satisfying
\begin{equation}\label{eq:asrhos}
    a_{s\downarrow}\equiv a(\eta_{s\downarrow})
  = \left(\frac{\rho_{c0}\,\Omega_{\gamma0}}{\rho_{s\downarrow}}\right)^{1/4}.
\end{equation}
Substituting this relation into Eq.~\eqref{eq:OmegaGWw1rhos} yields  
\begin{equation}\label{eq:OmegaGWw1as}
\begin{split}
  \Omega_{\rm GW}(f)\,h^2
  &= \frac{h^2}{24}\,
     \Bigl(\frac{f_{H_0}}{f_{m_{\rm pl}}}\Bigr)^2\,
     \frac{C(w_1)}{(2\pi)^{-n_T-1}}\,
     \Bigl(\frac{f}{f_{H_0}}\Bigr)^{n_T}\\
  &\quad\times
     \Bigl(\frac{a_{s\downarrow}}{\sqrt{\Omega_{\gamma0}}}\Bigr)^{-(4-n_T)}
     \,\mathcal{T}_{\rm eq}(f)\,,
\end{split}
\end{equation}
which is algebraically equivalent to Eq.~\eqref{eq:OmegaGWw1rhos} but expresses the spectrum in terms of the bounce scale factor \(a_{s\downarrow}\).

\begin{figure*}[htbp]
\centering 
\includegraphics[width=0.8\textwidth]{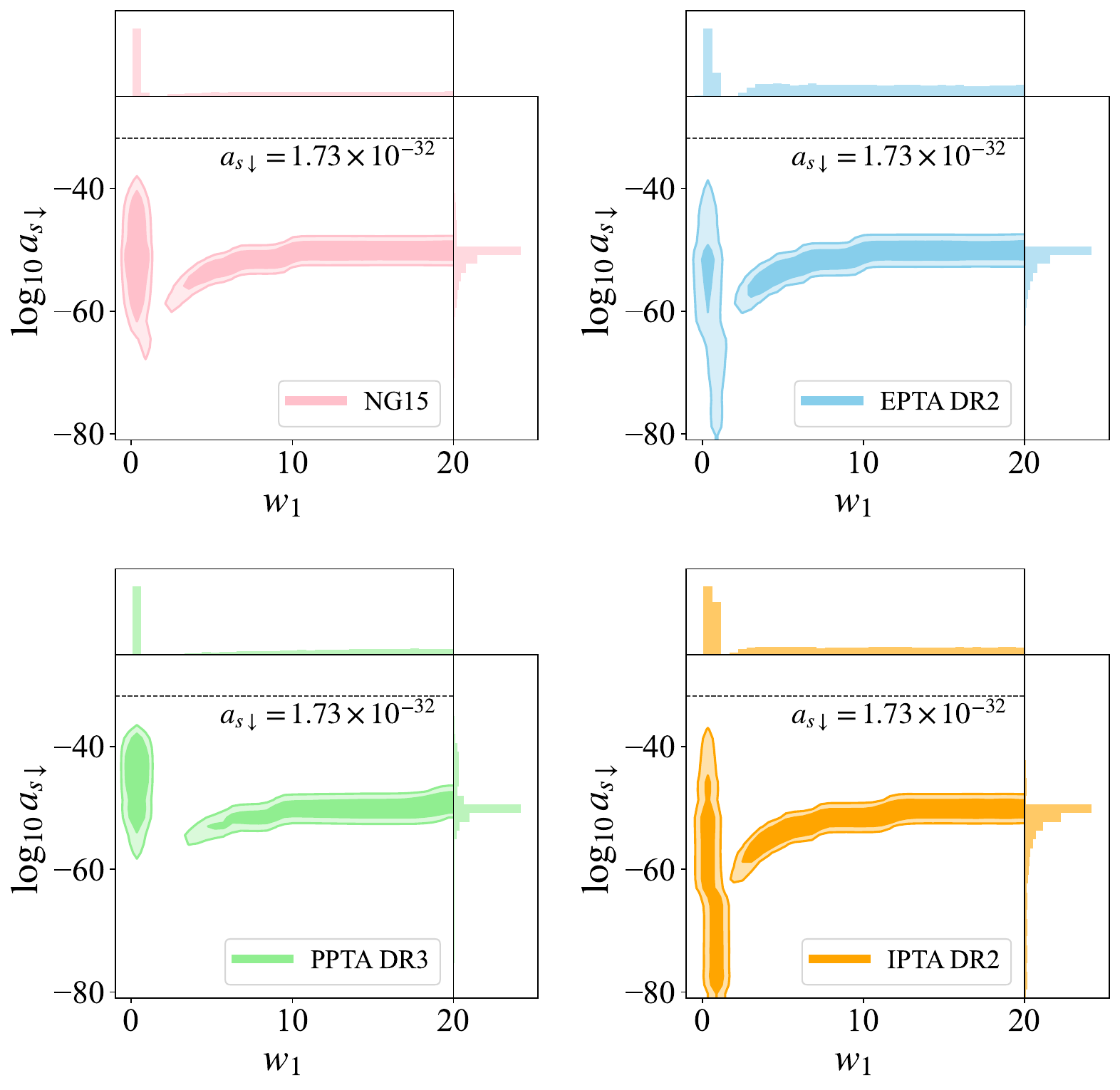}
\caption{\label{fig:2x2asw1}Two-dimensional posterior distribution in the \((a_{s\downarrow},w_1)\) plane, obtained by fitting Eq.~(\ref{eq:OmegaGWw1as}) to the combined PTA datasets (NANOGrav 15-yr, EPTA DR2, PPTA DR3, IPTA DR2).  The dashed horizontal line marks the  Planck scale factor $a_{\rm pl} \;=\;\bigl(\rho_{c0}\,\Omega_{\gamma0}/M_{\rm pl}^4\bigr)^{1/4}\simeq1.73\times10^{-32}\,$.}  

\end{figure*}

By performing a Bayesian fit of Eqs.~(\ref{eq:OmegaGWw1as}) and (\ref{eq:OmegaGWw1rhos}) to the full PTA datasets (NANOGrav 15-yr, EPTA DR2, PPTA DR3, IPTA DR2)~\cite{NANOGrav:2023gor,EPTA:2023fyk,Reardon:2023gzh,Antoniadis:2022pcn,ZenodoNG,NANOGrav:2023hde,ZenodoEPTA,EPTA:2023sfo,PPTADR3,Zic:2023gta,IPTADR2,Perera:2019sca}, we derive the two-dimensional posterior in the \((a_{s\downarrow},w_1)\) plane for a generic bouncing cosmology (Fig.~\ref{fig:2x2asw1}). These constraints on the bounce scale factor \(a_{s\downarrow}\) are equivalent—via Eq.~\eqref{eq:asrhos}—to those on \(\rho_{s\downarrow}^{1/4}\) shown in Fig.~\ref{fig:2x2rhosw1}, providing a dual characterization of the bounce energy scale.Akin to the constraints on \(\rho_{s\downarrow}^{1/4}\), the \((a_{s\downarrow},w_1)\) posterior (Fig.~\ref{fig:2x2asw1}) exhibits two notable features with implications for novel physics:  
(1) The “left” branch satisfies all standard energy conditions, whereas the “right” branch violates the dominant energy condition, thereby offering complementary empirical motivation to explore mechanisms—such as ghost condensates or modified-gravity effects—that permit energy‐condition violation during the very early Universe.  
(2) The best‐fit bounce scale factor \(a_{s\downarrow}\) lies below the Planck scale factor  
\begin{equation}
    a_{\rm pl}\;\equiv\;\left(\frac{\rho_{c0}\,\Omega_{\gamma0}}{M_{\rm pl}^4}\right)^{1/4}\!,
\end{equation}
indicating that PTA measurements probe length scales shorter than the Planck length in the bouncing cosmology.  

In Table~\ref{tab:a_vs_other_1f} we present the Bayes factors for the bouncing‐universe model \((a_{s\downarrow},w_1)\) versus six conventional SGWB sources.  Nearly all values exceed unity, indicating a generic preference for the bounce scenario.  These findings agree with the Bayes‐factor results for \((\rho_{s\downarrow}^{1/4},w_1)\) in Table~\ref{tab:rho_vs_other_1f}; the minor differences arise from the chosen prior range in \(a_{s\downarrow}\) used in the inference.  

\begin{table*}[htbp]
\centering
\caption{Bayes factors for the bouncing‐universe model \((a_{s\downarrow},w_1)\) versus six conventional SGWB sources}
\small 
\begin{tabular}{cccccccc}
\toprule
\textbf{Data Set} & \textbf{$(a_{s\downarrow},w_1)$/SMBHBs} & \textbf{$(a_{s\downarrow},w_1)$/IGW} & \textbf{$(a_{s\downarrow},w_1)$/CS} & \textbf{$(a_{s\downarrow},w_1)$/DW} & \textbf{$(a_{s\downarrow},w_1)$/FOPT} & \textbf{$(a_{s\downarrow},w_1)$/SIGW} \\ \hline
\textbf{NG15} & 25.9 ± 10.1 & 1.7 ± 0.7 & 126.5 ± 36.7 & 87.4 ± 21.2 & 16.8 ± 3.4 & 17.9 ± 3.6 \\
\textbf{EPTA DR2} & 1.4 ± 0.6 & 1.4 ± 0.5 & 1.1 ± 0.4 & 10.8 ± 2.8 & 1.2 ± 0.3 & 2.2 ± 0.5 \\ 
\textbf{PPTA DR3} & 5.4 ± 2.3 & 0.7 ± 0.4 & 3.9 ± 1.7 & 4.8 ± 2.2 & 1.5 ± 0.6 & 2.9 ± 1.0 \\
\textbf{IPTA DR2} & 0.9 ± 0.5 & 1.3 ± 0.4 & 1.9 ± 0.9 & 1.8 ± 1.0 & 0.7 ± 0.3 & 3.5 ± 1.7 \\ \bottomrule
\end{tabular}
\label{tab:a_vs_other_1f}
\end{table*}

\section{Summary and Prospects}

We have presented the first Bayesian determination of the bounce energy scale \(\rho_{s\downarrow}^{1/4}\) in a generic bouncing-universe model that can account for the common-spectrum process, possibly the first hint of a nanohertz SGWB, reported by PTAs (NANOGrav 15-yr, EPTA DR2, PPTA DR3, IPTA DR2).  Fitting the analytic SGWB spectrum derived in Ref.~\cite{Li:2024dce}, we uncover two viable posterior branches in \((\rho_{s\downarrow}^{1/4},w_1)\): one near \(w_1\approx0.3\) and one at \(w_1\gg1\).  

Against six standard SGWB sources (SMBHBs, IGW, CS, DW, FOPT, SIGW), the bounce model attains larger Bayes factors, demonstrating strong preference by PTA data.  Recasting these bounds into the dual \((a_{s\downarrow},w_1)\) parameterization yields consistent conclusions.  By contrast, comparison with the generic dual inflation–bounce scenario \((w,r)\) produces smaller Bayes factors, indicating that making the bounce scale explicit tightens PTA constraints.

These two posterior regions also illuminate two frontiers of early-Universe physics:  
\begin{enumerate}
  \item \textbf{Energy-condition violation.}  The “left” branch (\(w_1\approx0.3\)) satisfies all standard energy conditions, whereas the “right” branch (\(w_1\gg1\)) violates the dominant energy condition (DEC).  The latter thus provides direct empirical impetus to explore novel mechanisms—e.g.\ ghost condensates, higher-derivative or modified-gravity operators, extra dimensions—that permit DEC violation in the pre-bounce phase.  
  \item \textbf{Trans-Planckian sensitivity.}  Both branches infer \(\rho_{s\downarrow}^{1/4}>M_{\rm pl}\equiv G^{-1/2}\simeq1.22\times10^{16}\,\mathrm{TeV}\), demonstrating that current PTAs already probe trans-Planckian regimes.  Consequently, any UV completion must satisfy these observational bounds, offering a rare window into physics beyond the Planck scale.  
\end{enumerate}

Finally, we have shown (Sec.~\ref{sec:NFbasedML}) that a normalizing-flow–based machine-learning emulator reproduces our MCMC posterior at greatly reduced cost, paving the way for rapid Bayesian inference as PTA datasets—especially with future facilities such as SKA—continue to expand. Last but not least, in Sec.~\ref{sec:furtherconstraints} we discuss the detectability of the bouncing‐universe SGWB by space‐ and ground‐based GW searches at higher frequencies, as well as additional constraints from BBN and CMB bounds.\footnote{We thank the anonymous referee for this suggestion, which has illuminated directions for future work.} These considerations imply that a broken power‐law feature must be incorporated into the SGWB spectrum of bouncing cosmologies to satisfy BBN and CMB limits and to enable detection by current and forthcoming space‐ and ground‐based GW detectors at higher frequencies.

\begin{acknowledgements}
C.L. is supported by the NSFC under Grants No.11963005 and No. 11603018, by Yunnan Provincial Foundation under Grants No.202401AT070459, No.2019FY003005, and No.2016FD006, by Young and Middle-aged Academic and Technical Leaders in Yunnan Province Program, by Yunnan Provincial High level Talent Training Support Plan Youth Top Program, by Yunnan University Donglu Talent Young Scholar, and by the NSFC under Grant No.11847301 and by the Fundamental Research Funds for the Central Universities under Grant No. 2019CDJDWL0005.
\end{acknowledgements}

\appendix

\section{SGWB sources: model descriptions, priors and posterior distributions}
\label{sec:SGWBmodels}
\subsection{Model descriptions}
\label{sec:Mdescri}
In this section we summarize each SGWB source model, following the definitions and notation of Refs.~\cite{Li:2024oru,Lai:2025xov}.

\begin{enumerate}
  \item \textbf{Model-1:} Bouncing–universe model \((\rho_{s\downarrow}^{1/4},w_1)\), Spectrum in Eq.~\eqref{eq:OmegaGWw1rhos}, where $\rho_{s\downarrow}^{1/4}$ is bounce energy scale, and $w_1$ is equation-of-state parameter during Phase I.  
    
  \item \textbf{Model-2:} Bouncing–universe model \((a_{s\downarrow},w_1)\), Spectrum in Eq.~\eqref{eq:OmegaGWw1as}, where $a_{s\downarrow}$ is the bounce scale factor.

  \item \textbf{Model-3:} SMBHBs (supermassive-black-hole binaries) \cite{Hobbs:2009yn,Bian:2022qbh,Jaffe:2002rt,Wyithe:2002ep} 
    \begin{equation}\label{eq:omeSMBHBs}
      \Omega_{\rm GW}^{\rm SMBHB}h^2
      =A_{\rm SMBHB}^2\,\frac{2\pi^2}{3H_0^2}\,f^{5-\gamma}f_{\rm yr}^{\gamma-3}h^2,
      \quad\gamma=\tfrac{13}{3},
    \end{equation}  
    where $A_{\rm SMBHB}$ is characteristic strain and $f_{\rm yr}=1\,\mathrm{yr}^{-1}$ .

  \item \textbf{Model-4:} Inflationary GWs/dual scenario \((n_T,r)\) \cite{Caprini:2018mtu, Li:2024oru} 
    \begin{equation}\label{eq:omentw}
      \Omega_{\rm GW}h^2
      =\tfrac{3}{128}\,\Omega_{\gamma0}h^2\,r\,P_R\,(f/f_\ast)^{n_T}
      \bigl[(f_\mathrm{eq}/f)^2+16/9\bigr],
    \end{equation}  
    where $n_T$ is tensor tilt; $r$ tensor-to-scalar ratio; $P_R$ scalar power at pivot $f_\ast$; and $f_\mathrm{eq}$ matter–radiation equality frequency .  
  \item \textbf{Model-5:} Cosmic strings (CS) \cite{NANOGrav:2023hvm,Blanco-Pillado:2017oxo,Chang:2021afa,Auclair:2019wcv,Martins:2000cs} 
    \begin{equation} \label{eq:omeCS}
      \Omega_{\rm GW}^{\rm cs}h^2
      =\frac{8\pi}{3H_0^2}(G\mu)^2\sum_{k=1}^{k_{\max}}P_k\,\mathcal{I}_k(f)\,h^2,
      \quad P_k=\frac{\Gamma}{\zeta(q)\,k^q},
    \end{equation}  
    where $G\mu$ is string tension; $P_k$ the GW power emitted by a loop in its $k$-th excitation, $\zeta(q)$ a factor ensuring $\sum_k P_k = \Gamma = 50$, $q = 4 / 3$ for cusps-only GW emission, and $\mathcal{I}_k(f)$ the frequency-dependent factor.  
  \item \textbf{Model-6:} Domain walls (DW) \cite{Hiramatsu:2013qaa,Kadota:2015dza,Zhou:2020ojf} 
    \begin{subequations}\label{eq:omeDW}
    \begin{align}
      \Omega_{\rm GW}^{\rm DW}h^2
      &= \Omega_{\mathrm{GW}}^{\text{peak}}h^2 \, S^{\rm dw}(f),
        \label{eq:omeDW_Omega}\\
      S^{\rm dw}(f)
      &=
      \begin{cases}
        \bigl(f/f_{\rm peak}\bigr)^3,   & f < f_{\rm peak},\\
        \bigl(f/f_{\rm peak}\bigr)^{-1},& f \ge f_{\rm peak},
      \end{cases}
        \label{eq:omeDW_S}
    \end{align}
    \end{subequations}

    \begin{equation}
    \begin{split}
      \Omega_{\mathrm{GW}}^{\text{peak}} h^2
      &\simeq 5.20 \times 10^{-20}\,\tilde{\epsilon}_{\mathrm{gw}}\,\mathcal{A}^4
        \Bigl(\frac{10.75}{g_\ast}\Bigr)^{1/3}\\
      &\quad\times \Bigl(\frac{\sigma}{1\,\mathrm{TeV}^3}\Bigr)^4
        \Bigl(\frac{1\,\mathrm{MeV}^4}{\Delta V}\Bigr)^2
    \end{split}
    \end{equation}
    
    where $\Omega_{\mathrm{GW}}^{\text{peak}}$ is peak SGWB spectrum; $f_{\rm peak}$ peak frequency; $\sigma$ wall tension; $\Delta V$ vacuum bias; $\mathcal{A}=1.2$ DW area parameter, and $\tilde\epsilon_{\rm gw}=0.7$ efficiency factor. For recent development, see Refs.~\cite{Babichev:2023pbf,Dankovsky:2024ipq} 
  \item \textbf{Model-7:} First-order phase transition (FOPT) \cite{Caprini:2015zlo,Bian:2023dnv,Hirose:2023bvl,Zhou:2020ojf}  
    \begin{equation}\label{eq:omeFOPT}
    \begin{split}
      \Omega_{\rm GW}^{\rm FOPT}h^2
      &= 2.65\times10^{-6}\,(H_\ast\tau_{\rm sw})\,(\beta/H_\ast)^{-1}v_b\\
      &\quad\times \Bigl(\frac{\kappa_v\alpha_{PT}}{1+\alpha_{PT}}\Bigr)^2
        F\bigl(f/f_{\text{peak}}^{\mathrm{FOPT}}\bigr)\,,
    \end{split}
    \end{equation}
    with the peak frequency
    \begin{equation}
    f_{\text{peak}}^{\mathrm{FOPT}}=1.9 \times 10^{-5} \frac{\beta}{H_\ast} \frac{1}{v_b} \frac{T_\ast}{100}\left(\frac{g_\ast}{100}\right)^{\frac{1}{6}} \mathrm{Hz},
    \end{equation}
    where $H_\ast$ is Hubble at $T_\ast$; $\tau_{\rm sw}$ sound-wave duration; $\beta/H_\ast$ inverse transition duration; $v_b$ bubble velocity; $\kappa_v,\alpha_{PT}$ energy-partition parameters; $g_\ast$ d.o.f., and $F(x)=x^3[7/(4+3x^2)]^{7/2}$.  
  \item \textbf{Model-8:} Scalar-induced GWs (SIGW) \cite{Kohri:2018awv,Bian:2023dnv}  
    \begin{equation}\label{eq:omeSIGW}
    \begin{aligned}
    \Omega_{\rm GW}^{\rm SI}h^2
    &=\tfrac{1}{12}\,\Omega_{\rm rad}h^2\Bigl(\tfrac{g_0}{g_\ast}\Bigr)^{1/3}\\
    &\qquad\times\int\!du\,dv\,\mathcal{K}(u,v)\,
       P_{\mathcal{R}}(k u)\,P_{\mathcal{R}}(k v)\,.
    \end{aligned}
    \end{equation}
    with the kernel
    \begin{equation}
    \begin{aligned}
    \mathcal{K}(u,v) & =\left(\frac{4v^2-\bigl(1+v^2-u^2\bigr)^2}{4uv}\right)^2\\
    &\times
    \frac{1}{2}\!\left(\frac{3}{4u^3v^3x}\right)^{2}
    \,(u^2+v^2-3)^2\\
    & \times \left\{\left[-4 u v+\left(u^2+v^2-3\right) \ln \left|\frac{3-(u+v)^2}{3-(u-v)^2}\right|\right]^2\right. \\
    & \left.+\left[\pi\left(u^2+v^2-3\right) \Theta(u+v-\sqrt{3})\right]^2\right\} ,
    \end{aligned}
    \end{equation}
    and the power spectrum of amplified curvature perturbations, 
    \begin{equation}
    P_{\mathcal{R}}(k)=P_{\mathcal{R} 0}\left(\frac{k}{k_\star}\right)^n \Theta\left(k-k_{\min}\right) \Theta\left(k_{\max}-k\right),
    \end{equation}
    where $k=2\pi f$, $P_{\mathcal{R} 0}$ is the primordial power spectrum of scalar perturbations, $n$ spectral slope of amplified scalar perturbation, $k_{\min} = 0.03 k_\ast$ and $k_{\max} = 100 k_\ast$ are cutoffs, and $k_\star = 2 \pi f_\mathrm{yr}=20.6 \mathrm{pc}^{-1}$.

  \item \textbf{Model-9:} Dual scenario (inflation+bounce) \((w,r)\)  \cite{Li:2024oru}
    \begin{equation}
      \Omega_{\rm GW}h^2
      =\tfrac{3}{128}\,\Omega_{r0}h^2\,r\,P_R\,(f/f_\ast)^{\frac{4}{3w+1}+2}
      \bigl[(f_\mathrm{eq}/f)^2+16/9\bigr],
    \end{equation}  
    where $w$ is background EoS and $r,P_R,f_\ast,f_\mathrm{eq}$ as above.

  \item \textbf{Model-10:} Dual scenario (only bounce) \((w>1,r)\) \cite{Li:2024oru}. Same as Model-9 with prior restricted to \(w>1/3\).  
\end{enumerate}
\newpage

\subsection{Priors}  
Table~\ref{tab:priors} summarizes the prior ranges adopted for all SGWB source models in our Bayesian analyses.

\begin{table*}[!t]
\centering
\caption{Parameter descriptions of SGWB source models and their priors}
\label{tab:priors} 
\begin{tabular}{lccc}
\toprule[0.4mm]
\textbf{Model} & \textbf{Parameter} & \textbf{Description} & \textbf{Prior} \\
\midrule 

\multirow{2}{*}{\makecell[l]{\textbf{Model-1:} \\ $(\rho^{1/4}_{s\downarrow}, w_1)$}}
 & $\rho^{1/4}_{s\downarrow}\;[\text{TeV}]$ & Bounce energy scale & log-uniform $[15, 50]$ \\
 & $w_1$ & Equation of state & uniform $[-1/3, 20]$ \\
\midrule

\multirow{2}{*}{\makecell[l]{\textbf{Model-2:} \\ $(a_{s\downarrow}, w_1)$}}
 & $a_{s\downarrow}$ & Bounce scale factor & log-uniform $[-65, -30]$ \\
 & $w_1$ & Equation of state & uniform $[-1/3, 20]$ \\
\midrule

\multirow{1}{*}{\makecell[l]{\textbf{Model-3: SMBHBs}}} 
 & $A$ & Amplitude of the signal & log-uniform $[-18, -12]$\\
\midrule

\multirow{2}{*}{\makecell[l]{\textbf{Model-4: IGW} \\ $(n_T, r)$}}
 & $n_T$ & Spectral index of tensor spectrum & uniform $[0, 6]$ \\
 & $r$ & Tensor-to-scalar ratio & log-uniform $[-16, 0]$ \\
\midrule

\multirow{1}{*}{\makecell[l]{\textbf{Model-5: CS}}}
 & $G\mu$ & Cosmic-string tension & log-uniform $[-12, -6]$\\
\midrule

\multirow{2}{*}{\makecell[l]{\textbf{Model-6: DW}}}
 & $\sigma$ & Surface energy density & log-uniform $[0, 8]$\\
 & $\Delta V$ & Bias potential & log-uniform $[0, 8]$\\
\midrule

\multirow{4}{*}{\makecell[l]{\textbf{Model-7: FOPT}}}
 & $\beta /H_\star$ & Inverse PT duration & uniform $[5, 70]$\\
 & $T_\star [\mathrm{MeV}]$ & PT temperature & uniform $[0.01, 1.6]$\\
 & $\alpha_{PT}$ & PT strength & uniform $[0.0, 1.0]$\\
 & $\nu_b$ & Velocity of bubble wall & uniform $[0.9, 1.0]$\\
\midrule

\multirow{2}{*}{\makecell[l]{\textbf{Model-8: SIGW}}}
 & $P_{R0}$ & Amplitude of signal & log-uniform $[-4, -1.5]$\\
 & $n$ & Slope of spectrum & uniform $[-2, 2]$\\
\midrule

\multirow{2}{*}{\makecell[l]{\textbf{Model-9: Dual scenario} \\ $(w, r)$}}
 & $w$ & Equation of state & uniform $[-10, 10]$ \\
 & $r$ & Tensor-to-scalar ratio & log-uniform $[-16, 0]$ \\
\midrule

\multirow{2}{*}{\makecell[l]{\textbf{Model-10: Bounce} \\ $(w, r)$}}
 & $w$ & Equation of state & uniform $[1/3, 10]$ \\
 & $r$ & Tensor-to-scalar ratio & log-uniform $[-16, 0]$ \\

\bottomrule[0.4mm] 
\end{tabular}
\end{table*}

\subsection{Posteriors}  
\begin{figure}[htbp]
\centering 
\includegraphics[width=0.5\textwidth]{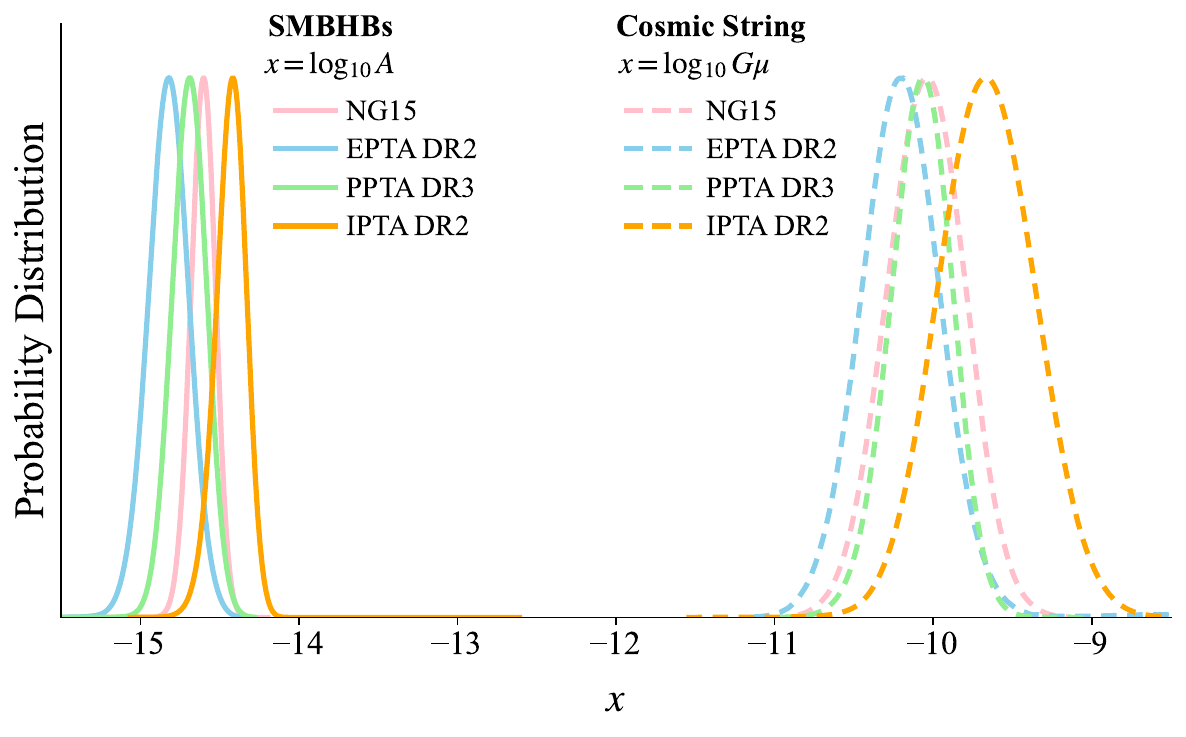}
\caption{\label{fig:SMBHBs_CS}One-dimensional posterior distributions (probability distribution) for the SMBHB model (solid line) and the cosmic-string model (dashed line), obtained from Bayesian fits of Eqs.~\eqref{eq:omeSMBHBs} and \eqref{eq:omeCS} to the full PTA dataset.}
\end{figure}

\begin{figure}[htbp]
\centering 
\includegraphics[width=0.5\textwidth]{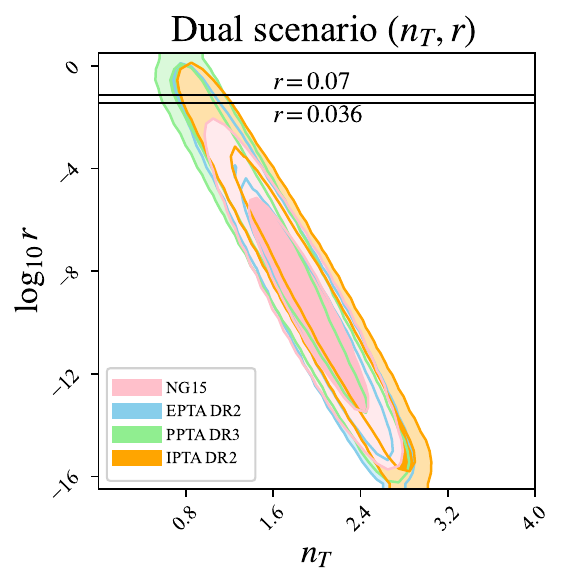}
\caption{\label{fig:n_Tr}Two-dimensional posterior distribution for the inflationary GWs/dual scenario \((n_T,r)\) model, obtained from a Bayesian fit of Eq.~\eqref{eq:omentw} to the full PTA dataset.}
\end{figure}

\begin{figure}[htbp]
\centering 
\includegraphics[width=0.5\textwidth]{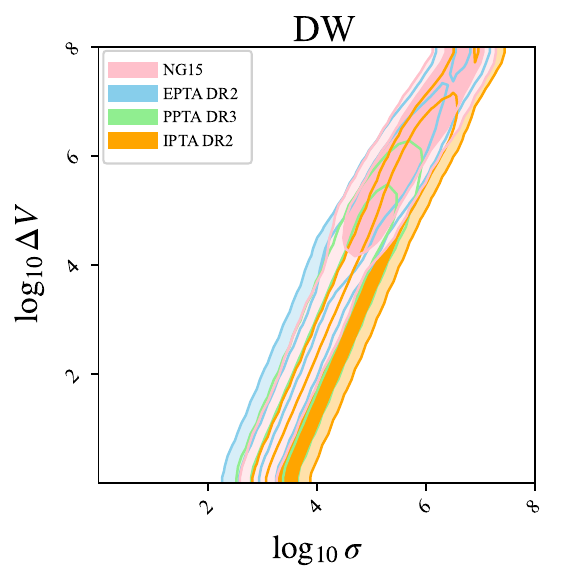}
\caption{\label{fig:DW}Two-dimensional posterior distribution for the domain-wall model, obtained from a Bayesian fit of Eq.~\eqref{eq:omeDW} to the full PTA dataset.}
\end{figure}

\begin{figure}[htbp]
\centering 
\includegraphics[width=0.5\textwidth]{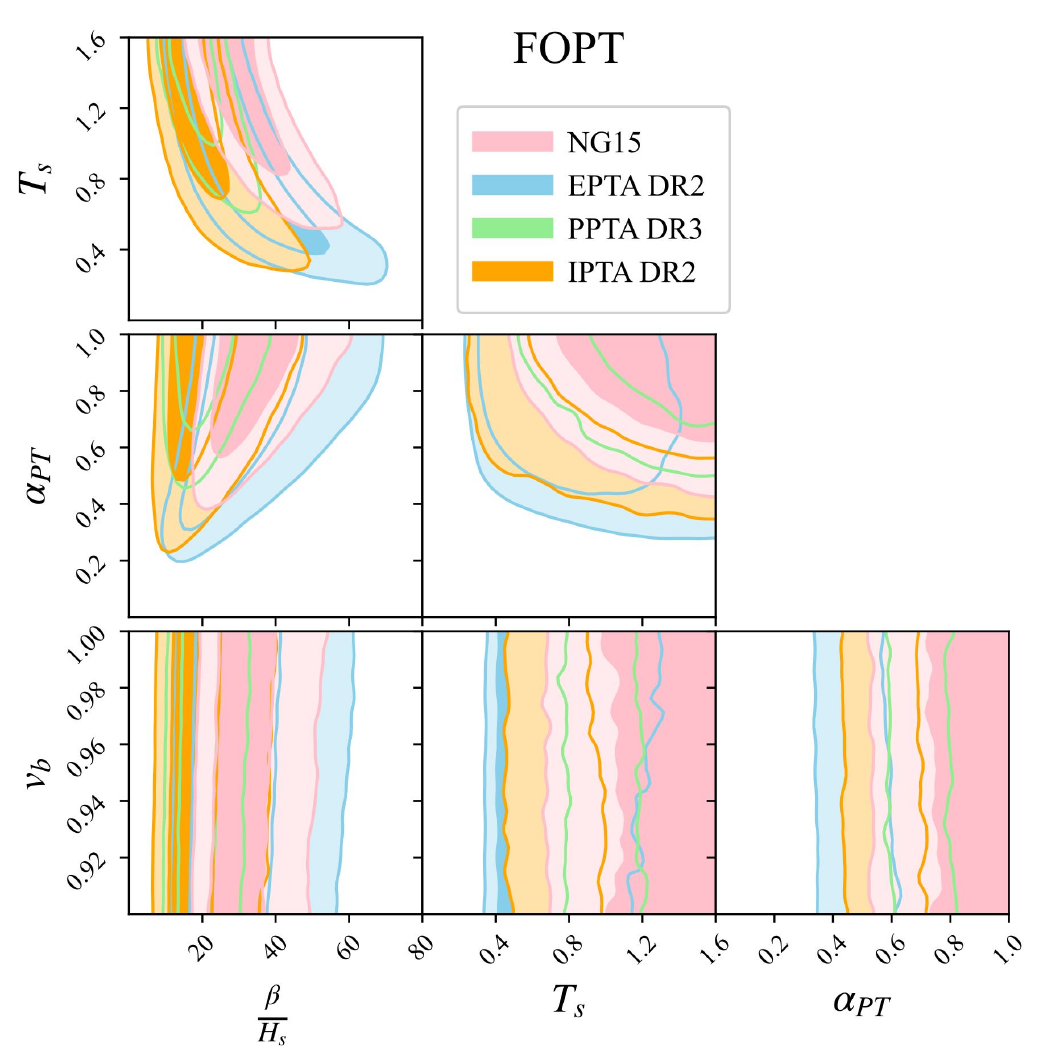}
\caption{\label{fig:FOPT}Two-dimensional posterior distribution for the domain-wall model, obtained from a Bayesian fit of Eq.~\eqref{eq:omeFOPT} to the full PTA dataset.}
\end{figure}

\begin{figure}[htbp]
\centering 
\includegraphics[width=0.5\textwidth]{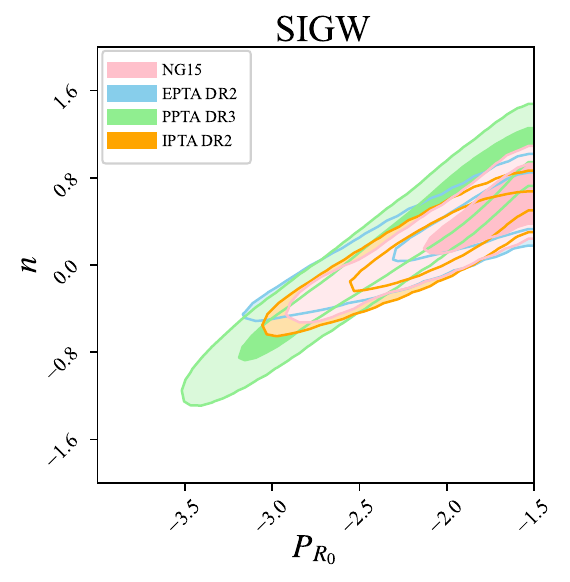}
\caption{\label{fig:SIGW}Two-dimensional posterior distribution for the SIGW model, obtained from a Bayesian fit of Eq.~\eqref{eq:omeSIGW} to the full PTA dataset.}
\end{figure}

Figures~\ref{fig:SMBHBs_CS}–\ref{fig:SIGW} display the one- and two-dimensional posterior distributions for the six conventional SGWB source models: supermassive-black-hole binaries (SMBHBs), inflationary gravitational waves (IGW), cosmic strings (CS), domain walls (DW), first-order phase transitions (FOPT), and scalar-induced GWs (SIGW), obtained by employing \texttt{Ceffyl}~\cite{Lamb:2023jls} to perform Bayesian fit  of each model (Sec.~\ref{sec:Mdescri}) to the full PTA dataset ( NANOGrav 15-year, EPTA DR2, PPTA DR3 and IPTA DR2)~\cite{NANOGrav:2023gor,EPTA:2023fyk,Reardon:2023gzh,Antoniadis:2022pcn,ZenodoNG,NANOGrav:2023hde,ZenodoEPTA, EPTA:2023sfo,PPTADR3,Zic:2023gta, IPTADR2,Perera:2019sca}, following Ref.~\cite{Li:2024oru} (c.f. Fig.2 and Fig.7-10 of \cite{Li:2024oru}).  

\section{Accelerating Bayesian Inference with Normalizing-Flow Machine Learning}
\label{sec:NFbasedML}

\begin{figure*}[htbp]
\centering 
\includegraphics[width=0.8\textwidth]{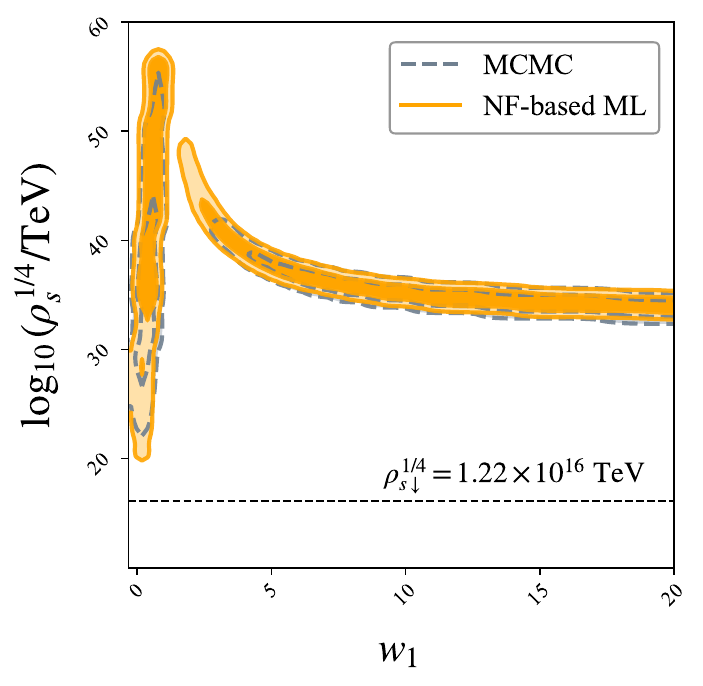}
\caption{\label{fig:NFML}Comparison of posterior distributions in the \((\rho_{s\downarrow}^{1/4},w_1)\) plane inferred via MCMC (blue dashed contours) and via a normalizing‐flow–based machine‐learning emulator (filled shading), trained on the ten most sensitive pulsars from the NANOGrav 15-yr dataset~\cite{ZenodoNG}.  Both methods fit Eq.~(\ref{eq:OmegaGWw1rhos}).  The dashed horizontal line denotes the Planck scale \(M_{\rm pl}= G^{-1/2}\simeq1.22\times10^{16}\,\mathrm{TeV}\).}
\end{figure*}

\begin{figure*}[htbp]
\centering 
\includegraphics[width=0.8\textwidth]{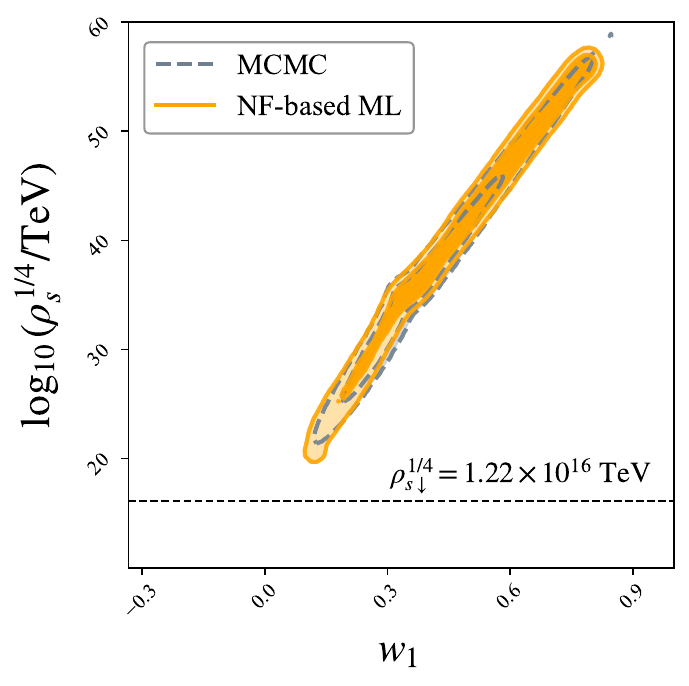}
\caption{\label{fig:NFML131}High‐resolution view of the “left” posterior branch (\(w_1\sim0.3\)) from Fig.~\ref{fig:NFML}, highlighting the consistency between MCMC (blue dashed) and normalizing‐flow–based machine‐learning (shaded) inferences.}
\end{figure*}

In Fig.~\ref{fig:NFML} we show the posterior in the \((\rho_{s\downarrow}^{1/4},w_1)\) plane obtained from a normalizing-flow–based machine-learning architecture trained on the ten most sensitive pulsars from the NANOGrav 15-yr dataset~\cite{ZenodoNG}.  Following the methodology of Ref.~\cite{Lai:2025xov} (see also Refs.~\cite{Shih:2023jme,Vallisneri:2024xfk}), the flow is trained on the analytic SGWB spectrum of the bouncing model (Eq.~\ref{eq:OmegaGWw1rhos}). The normalizing‐flow posterior closely reproduces our MCMC result (Fig.~\ref{fig:2x2rhosw1}, Sec.~\ref{sec:interpretation}) while reducing inference time from \(\sim10\) days to \(\sim20\) hours (see Ref.~\cite{Lai:2025xov} for technical details), demonstrating a scalable route to rapid Bayesian analyses as PTA datasets—and future arrays such as SKA—expand. Fig.~\ref{fig:NFML131} presents a high-resolution view of the “left” branch (\(w_1\sim0.3\)), highlighting the consistency between MCMC and the normalizing-flow–based approach.

\section{A Broken Power‐Law Feature for Higher Frequency GW Searches and BBN and CMB Constraints}
\label{sec:furtherconstraints}

Besides PTAs, the bouncing‐universe SGWB may also be probed by space‐ and ground‐based detectors (e.g., LISA, Taiji, Tianqin, aLIGO + aVirgo + KAGRA,  Einstein Telescope, Cosmic Explorer, see~\cite{Schmitz:2020syl, Annis:2022xgg, Bi:2023tib} for the summary of their sensitivities, and for recent developments in gravitational wave cosmology incorporating these GW detectors, for example, see~\cite{Bian:2025ifp, Yao:2025wfd, Chen:2023zkb, Chen:2024mwg} and reference therein), while BBN and CMB observations impose bounds on its amplitude over a broad frequency range.

\begin{figure*}[htbp]
\hspace*{-2cm}
\includegraphics[width=1.2\textwidth]{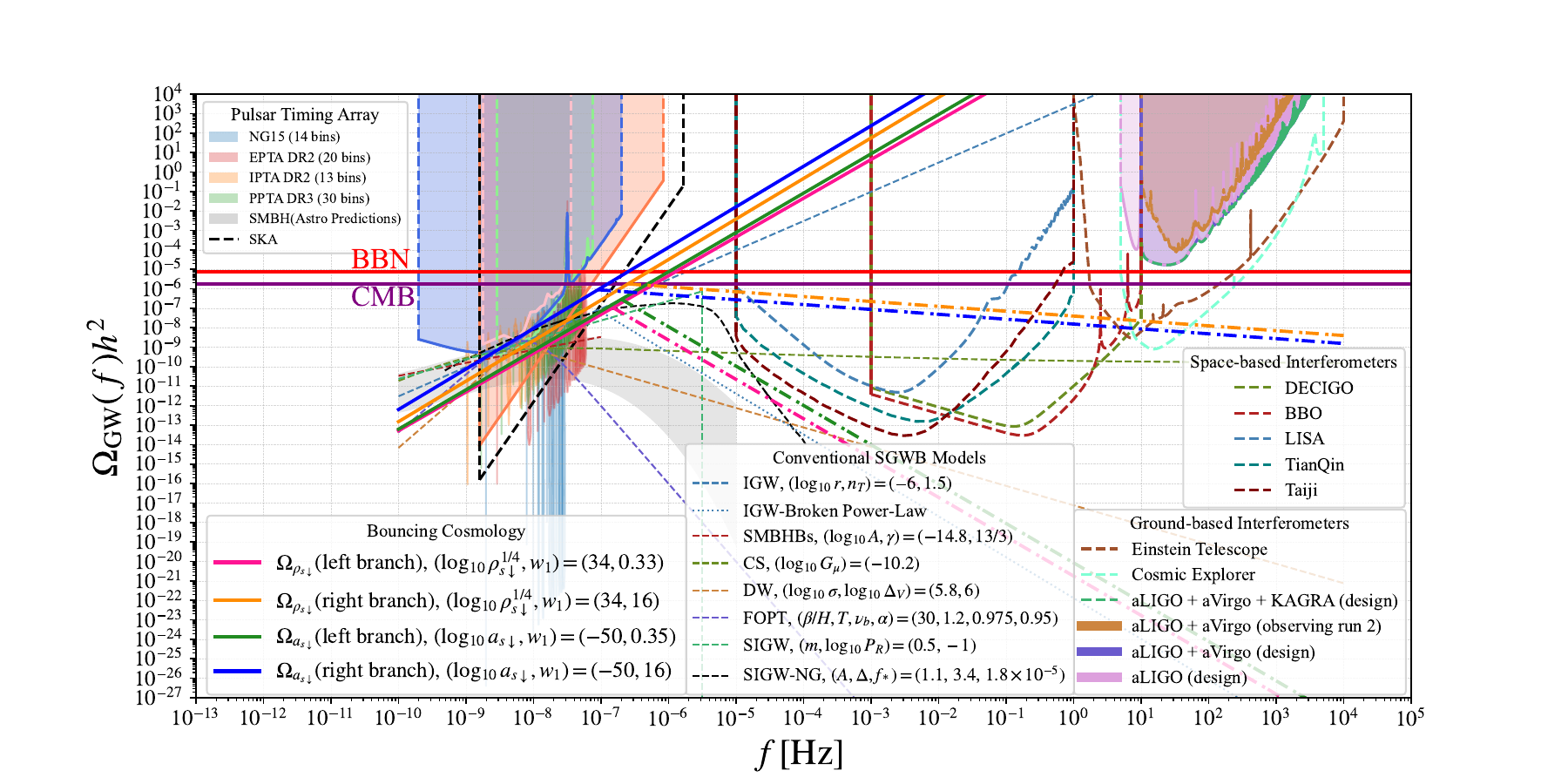}
\caption{\label{fig:OmegaGWfall} 
Forecast projections and sensitivities from current and forthcoming GW detectors (colored solid/dashed curves); BBN and CMB bounds (red/purple solid horizontal lines); and four representative examples of the bouncing‐universe model (solid pink, orange, green, and blue lines). Without a broken power‐law feature, these examples exceed the BBN and CMB limits for \(f>10^{-7}\)\,Hz. Consequently, the bouncing‐universe interpretation likewise requires a (quasi) broken power‐law feature (colored dash‐dotted lines) at frequencies just above the PTA band to remain consistent with BBN and CMB constraints, analogous to the inflationary interpretation (blue dashed line) and its broken power‐law feature (blue dotted line). The thin colored dashed curves correspond to six conventional SGWB models: astrophysical SMBHBs (dark red dashed line and gray band), cosmic strings (green), domain walls (brown), first‐order phase transitions (blue), SIGWs (sea‐green), and non‐Gaussianity (SIGW-NG, black). These models also exhibit (quasi) broken power‐law behavior. Furthermore, the prospect of probing a bouncing cosmology with space- and ground-based detectors (e.g., LISA, Taiji, Tianqin, aLIGO+aVirgo+KAGRA, Einstein Telescope, Cosmic Explorer) depends on the SGWB spectral index above the broken power‐law scale. For example, the orange and blue dash-dotted cases—with a mild red tilt after the break (\(n_{T}\sim0.25\) for a break at \(f\sim10^{-7}\)\,Hz)—lie within detector sensitivity, whereas the pink and green dash-dotted cases—with a steep red tilt (\(n_{T}<-2\) for a break at \(f\sim10^{-7}\)\,Hz)—fall below sensitivity at higher frequencies.
}
\end{figure*}

In Fig.~\ref{fig:OmegaGWfall}, we illustrate forecast projections and sensitivities from current and forthcoming GW detectors (colored solid and dashed curves, following~\cite{Schmitz:2020syl, Bi:2023tib}); BBN and CMB bounds (red and purple solid horizontal lines, $\Omega_{\mathrm{GW}}^{\mathrm{BBN}}h^2(f) \lesssim 7.8 \times 10^{-6}$ and $\Omega_{\mathrm{GW}}^{\mathrm{CMB}}h^2(f) \lesssim 1.7 \times 10^{-6}$, following \cite{Smith:2006nka, Pagano:2015hma} 
); and four representative examples of the bouncing‐universe model (solid pink, orange, green, and blue lines, following Fig.~\ref{fig:Violinplot}.). Without a broken power‐law feature, these examples exceed the BBN and CMB limits for \(f>10^{-7}\,\mathrm{Hz}\). This implies that the bouncing‐universe interpretation likewise requires a (quasi) broken power‐law feature (dash‐dotted pink, orange, green, and blue lines) at frequencies just above the PTA band to remain consistent with BBN and CMB constraints, analogous to the inflationary interpretation (blue dashed line) and its broken power‐law feature (blue dotted line)~\cite{Kuroyanagi:2020sfw,Benetti:2021uea}. The thin colored dashed curves correspond to six conventional SGWB models: astrophysical SMBHBs (dark red dashed line and gray band~\cite{Rosado:2015epa}), cosmic strings (green~\cite{Bian:2023dnv}), domain walls (brown~\cite{Bian:2022qbh}), first‐order phase transitions (blue~\cite{Bian:2023dnv}), SIGWs (sea‐green~\cite{Bian:2023dnv, Li:2024lxx}), and non‐Gaussianity (SIGW-NG, black~\cite{Liu:2023ymk}). These models also incorporate a (quasi) broken power‐law feature.

Furthermore, Fig.~\ref{fig:OmegaGWfall} also illustrates that the prospect of probing a bouncing cosmology with space‐ and ground‐based gravitational‐wave detectors (such as LISA, Taiji, Tianqin, etc.) depends on the SGWB spectral index above the broken power‐law scale. For example, the orange and blue dash‐dotted curves—with only a mild red tilt after the break (e.g., \(n_T \sim 0.25\) for a break at \(f \sim 10^{-7}\)Hz)—remain within the sensitivity of these detectors at higher frequencies, whereas the pink and green dash‐dotted curves—with a steep red tilt—fall below their sensitivity (e.g., \(n_T < -2\) for a break at \(f \sim 10^{-7}\)Hz).

\bibliography{biblio}
\bibliographystyle{spphys}

\end{document}